\documentclass[fleqn,10pt]{wlscirep}
\usepackage{color,soul}
\usepackage[utf8]{inputenc}
\usepackage[T1]{fontenc}
\usepackage{ulem}
\usepackage{subfigure}

\title{Oil spill risk analysis for the NEOM shoreline}

\author[1,3]{H.V.R. Mittal}
\author[2]{Mohamad Abed El Rahman Hammoud}
\author[1]{Ana K. Carrasco}
\author[2]{Ibrahim Hoteit}
\author[1]{Omar M. Knio*}
\affil[1]{Computer, Electrical and Mathematical Sciences and Engineering Division, King Abdullah University of Science and Technology, (KAUST), Thuwal 23955-6900, Saudi Arabia}
\affil[2]{Physical Science and Engineering Division, King Abdullah University of Science and Technology, (KAUST), Thuwal 23955-6900, Saudi Arabia}
\affil[3]{Department of Mathematics, Indian Institute of Technology Ropar, Rupnagar, Punjab 140001, India}

\affil[*]{omar.knio@kaust.edu.sa}

\keywords{The Red Sea, NEOM, Oil Spill, Risk Analysis, MOHID}

\begin{abstract}
A risk analysis is conducted considering an array of release sources located around the NEOM shoreline. The sources are selected close to the coast and in neighboring regions of high marine traffic.  The evolution of oil spills released by these sources is simulated using the MOHID model, driven by validated, high-resolution met-ocean fields of the Red Sea. For each source, simulations are conducted over a 4-week period, starting from first, tenth and twentieth days of each month, covering five consecutive years. A total of 48 simulations are thus conducted for each source location, adequately reflecting the variability of met-ocean conditions in the region. The risk associated with each source is described in terms of amount of oil beached, and by the elapsed time required for the spilled oil to reach the NEOM coast, extending from the Gulf of Aqaba in the North to Duba in the South. To further characterize the impact of individual sources, a finer analysis is performed by segmenting the NEOM shoreline,  based on important coastal development and installation sites.  For each subregion, source and release event considered, a histogram of the amount of volume beached is generated, also classifying individual events in terms of the corresponding arrival times.  In addition, for each subregion considered, an inverse analysis is conducted to identify regions of dependence of the cumulative risk, estimated using the collection of all sources and events considered.  The transport of oil around the NEOM shorelines is promoted by chaotic circulations and northwest winds in summer, and a dominant  cyclonic eddy in winter. Hence, spills originating from release sources located close to the NEOM shorelines are characterized by large monthly variations in arrival times, ranging from less than a week to more than two weeks. Similarly, large variations in the volume fraction of beached oil, ranging from less then 50\% to more than 80\% are reported.
The results of this study provide key information regarding the location of dominant oil spill risk sources, the severity of the potential release events, as well as the time frames within which mitigation actions may need to deployed.
\end{abstract}

\begin{document}

\flushbottom
\maketitle
\thispagestyle{empty}

% !TEX root = ./main_merged.tex

\section*{Introduction}
\begin{figure}[p]
    \centering
    \vspace{-1cm}
    %\hspace{-2.5cm}
    % \begin{subfigure}[]{}
    % \centering
    % \includegraphics[width=0.8\linewidth]{neom_topo.png}
    % \end{subfigure} \\
    % \begin{subfigure}[]{}
    \includegraphics[width=0.5\linewidth]{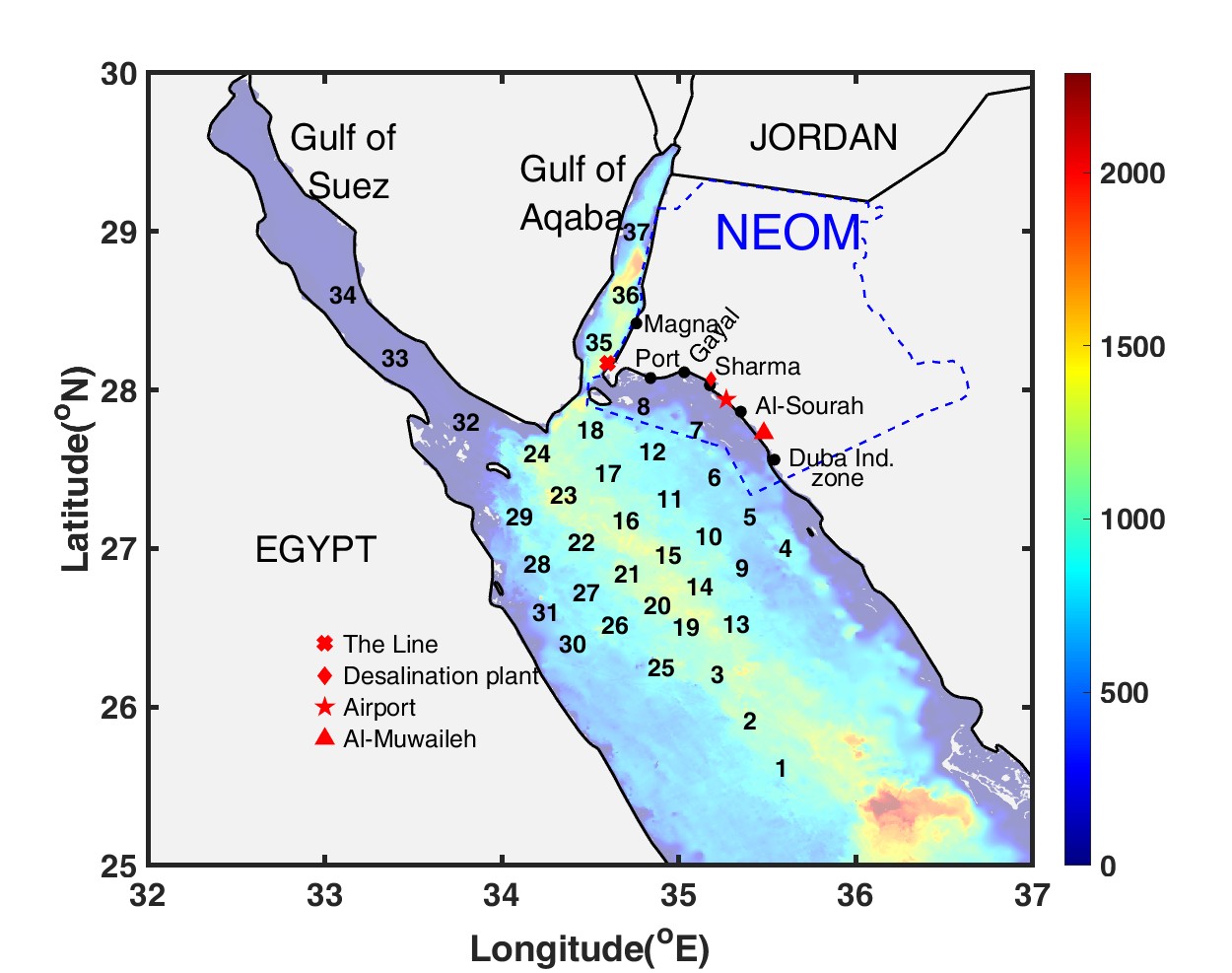}
    %     \end{subfigure}
    % \begin{subfigure}[]{}
    \includegraphics[width=0.41\linewidth]{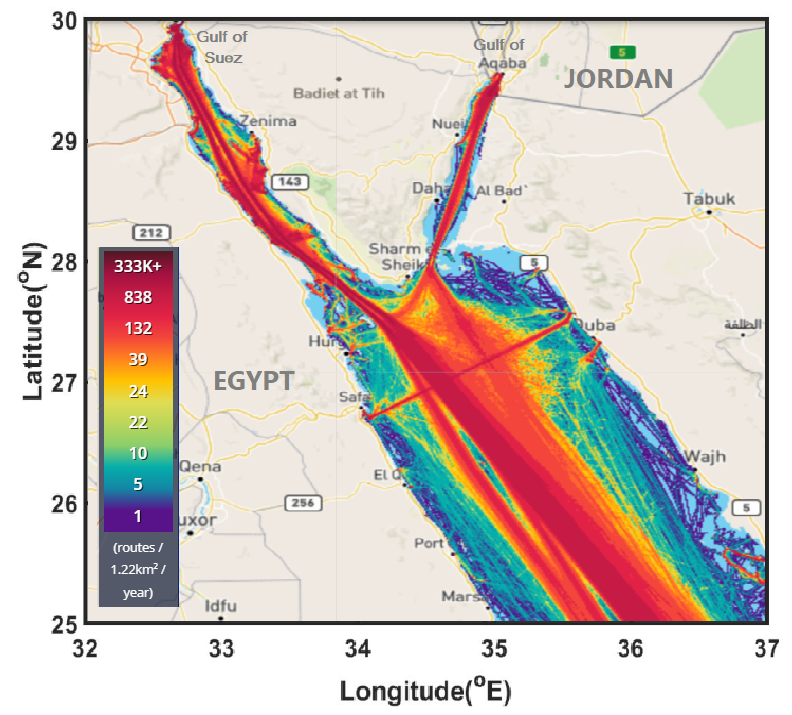}
    % \end{subfigure}
    \caption{(Left) General map of the northern Red Sea region illustrating key installations along the NEOM coastlines as well as the source locations chosen for spill simulations as well as the bathymetry of the northern Red Sea colored by depth in meters, and (Right) Contours exhibiting the marine traffic density, as the number of ships per 23 $km^2$ averaged over a year.\cite{noauthor_marine_traffic}}
    \label{schematic}
\end{figure}

NEOM is a smart city being developed in the Tabuk province\cite{noauthor_neom_nodate}, Kingdom of Saudi Arabia. It is situated in the north western part of the Kingdom with miles of Red Sea coastlines.  At its northernmost point, it is just 50 kilometers from the Jordanian port of Aqaba. NEOM development plans include establishing modern manufacturing facilities, industrial research and development, in addition to a hydrogen plant, a desalination plant and an international airport (see Fig.~\ref{schematic}). Tourism facilities are also being developed along its coastal environment hosting a diverse marine wildlife and coral reserves\cite{noauthor_neom_nodate}.

With an estimated 6.2 million barrels per day of crude oil and refined petroleum products transported through its main shipping lanes in 2018\cite{noauthor_EIA}, the Red Sea is one of the most active waterways in the world\cite{mittal2021hazard}. This poses a risk of accidental oil spills that may contribute to marine pollution, disrupting desalination operations, and consequently causing severe economic losses and irreversible damages to the environment \cite{kleinhaus2020closing, liu2009economy, mittal2021hazard, huynh2021public}. Therefore a comprehensive analysis of risk from accidental oil spill releases on coastal Red Sea regions is of paramount importance, particularly to minimize potential impact to both the environment and industrial activities, and to plan emergency response and mitigation efforts in case of an accident. 

Several studies assessed the risk of oil spill accidents for different regions around the world. These encompassed the Mediterranean sea\cite{alves2014three, alves2015modelling, alves2016multidisciplinary, olita2012oil, al2017risk}, the southern Adriatic and the northern Ionian (SANI) sea\cite{liubartseva2015oil}, Canadian waters\cite{marty2014risk}, Caribbean sea\cite{singh2015potential}, Sicily coasts \cite{canu2015assessment} and Bay of Bengal\cite{kankara2016environmental}.
A few studies have investigated the risk of oil spills on specific regions of the Red Sea, namely pertinent to the Egyptian coastlines \cite{nasr2006simulation}, the Bashayer shorelines \cite{ahmed2012modeling} and the Saudi Arabian-Yemeni coastlines \cite{huynh2021public}. Peri{\'a}{\~n}ez\cite{perianez2020lagrangian} presented a Lagrangian model for the whole Red Sea. Mittal et al.\cite{mittal2021hazard} provided a broad assessment of oil spill hazards for the whole Red Sea, stemming from its main shipping lane along the longitudinal axis. Pertinent to the risk analysis of oil spills for the NEOM shoreline, a study is still lacking, where existing studies in the literature that focus on NEOM encompass atmospheric conditions and air quality assessment\cite{dasari2020atmospheric}, geological assessment \cite{kahal2020geological, mogren2021geo} and wind energy assessment \cite{alfawzan2020wind} only. 

This study is part of an effort aimed at developing a fundamental understanding of the risk associated by possible oil release sources on the NEOM coastline, and consequently establishing a knowledge base that can assist in the design of efficient strategies safeguard its coastal environment from accidental oil spills.  Specifically, a hazard analysis is conducted considering an array of 37 potential release sources located around the NEOM coastline in regions of high marine traffic (see Fig.~\ref{schematic}). The risk associated with each source is described by the amount of oil beached following the initial release, and by the elapsed time required for the spilled oil to reach the NEOM coast. The evolution of the oil spill is simulated using the MOHID oil spill model \cite{noauthor_mohid_nodate,leitao2013overview, janeiro2008wind, mateus2008modelling}.  The model enables realistic, three-dimensional simulations of oil trajectories, accounting for weathering phenomena such as evaporation, dispersion, sedimentation, dissolution, and emulsification. Extensively-validated, high-resolution met-ocean fields\cite{hoteit2021towards} of the Red Sea are used to drive the oil spill model. For each release source, simulations are conducted over a 28-day period, starting from the first, tenth and twentieth days of each month, covering five consecutive years ranging from 2013 to 2017.  A total of 180 simulations are thus conducted for each source, adequately reflecting the variability of met-ocean conditions in the region.  In addition to characterizing the impact of individual sources, the simulation results are analyzed by segmenting the NEOM shoreline, extending from the Gulf of Aqaba in the North to Duba in the South, based on important coastal developments and installations.  For each subregion, an inverse analysis is finally conducted to identify regions of dependence of the cumulative risk estimated using the collection of sources considered.

%The study is structured as follows. The Methods and Data section details the met-ocean reanalysis of the Red Sea, the oil spill model employed for the simulations, followed by a brief description of the parameters used to quantify the  risk from individual release sources. The Results and Discussion section presents the simulation results in terms of the impacts on the entire NEOM shorelines followed by a finer analysis of smaller segments around specific sites (The Line, Duba, Magna, Sharma and Gayal). Key observations are summarized in the Conclusions section.

% !TEX root = ./main_merged.tex

\section*{Methods and Data}

\subsection*{Red Sea Met-Ocean Reanalysis}
\label{sec:methods}

Met-ocean data are extracted from an extensively-validated reanalysis of the circulation in the Red Sea \cite{hoteit2021towards}. The simulated fields have been shown to suitably describe the general oceanic and atmospheric circulations of the Red Sea at the highest available resolution \cite{hoteit2021towards, langodan2015wind, viswanadhapalli2017climatic, dasari2019high}.
The zonal and meridional winds were fetched from a 5 km regional atmospheric reanalysis generated using the Weather Research Forecasting (WRF) model assimilating all available regional observations \cite{viswanadhapalli2017climatic, dasari2019high}. WRF initial and boundary conditions were acquired from the European Centre for Medium-Range Weather Forecasts (ECMWF) reanalysis Interim data \cite{Dee2011ERA} (ERA-I). 
The wave conditions \cite{langodan2017climatology2} in the Red Sea were reconstructed using the WAVEWATCH III (WWIII) model forced with the aforementioned high-resolution WRF reanalysis winds \cite{langodan2017climatology} on a uniform grid of 1 km resolution. 

The MIT general circulation model (MITgcm \cite{marshall1997hydrostatic}) was implemented to simulate the 3D ocean currents on a grid with 1-km resolution in horizontal planes and $50$ vertical layers.
The model was forced using the aforementioned high-resolution WRF reanalysis fields and the Copernicus Marine Service Environment Monitoring Service (CMEMS) global ocean reanalysis fields \cite{noauthor_home_nodate} across the open-boundary in the Gulf of Aden at a 6 hourly and 24 hourly temporal frequency, respectively. 
The resulting MITgcm outputs for the Red Sea have been extensively employed to analyze the general and over-turning circulations \cite{yao2014seasonalb, yao2014seasonala}, internal/baroclinic tides \cite{guo2018baroclinic}, mesoscale eddies characteristics \cite{zhan2019three}, deep-water formation events \cite{yao2018rapid}, temperature and salinity budgets \cite{Krokos2022} as well as the chlorophyll variability \cite{gittings2018impacts}.
We refer readers to \cite{hoteit2021towards} for a more detailed description of the met-ocean conditions.

\subsection*{Northern Red Sea Circulation}

Mesoscale eddies \cite{zhan2016eddy, zhan2015far} play a dominant role in pollutant transport in the northern Red Sea region. A typical cyclonic eddy dominates the circulation during the winter season, and is characterized by a rotational velocity that are generally larger than that of the background flow \cite{mittal2021hazard}.
These eddies tend to become more energetic during winter months following the development of intense baroclinic instabilities \cite{zhan2014eddies, zhan2016eddy}, and they represent the dominant structures except for some strong semi-permanent wind-driven 
gyres that occur in summer \cite{zhai2013response}.

The high mountain ranges on both sides of the Red Sea forces the wind to blow along its axis \cite{langodan2014red}. 
During summer seasons, from April till October, a northwest (NW) wind blows along the whole length of The Red sea, with speeds close to  10 ms$^{-1}$, and frequently exceeding  15 ms$^{-1}$ \cite{langodan2017climatology}.
During winter, the same northerly wind dominates over the northern part of the basin. 
The narrower valleys along the eastern coasts of the Red Sea also creates westward blowing jets in the northern part and generally lasts for 2-3 days with a maximum speed up to 15 ms$^{-1}$. 
The wave variability in the Red Sea is naturally associated with the dominant regional wind regimes \cite{langodan2014red}. Despite the moderate winds, the prolonged duration and long fetch along the whole basin may generate waves as high as 3.5 m.  During the summer months, the northwesterly winds prevailing over the whole Red Sea generate mean wave heights of 1 m-1.5 m in the north \cite{langodan2014red, langodan2018unraveling}, throughout the year.

\subsection*{Oil Spill Model}

The MOHID oil spill model was adopted to simulate the instantaneous release of oil and its evolution from fixed sources 
in the northern Red Sea.  It relies on a Lagrangian formalism that considers oil as a collection of Lagrangian particles and associates to each particle oil properties and a location \cite{Batchelder2006, van_Sebille_2015}.
The Lagrangian particles are transported using the met-ocean conditions, and their properties are updated by solving empirical models describing physio-chemical transformations of oil.  Typically, these weathering processes result in changes in oil's physical properties and also impact the oil slick's geometry.  In the present study, dissolution and sedimentation processes were not considered, thus eliminating their effect on the oil mass balance.  However, evaporation, dispersion and emulsification were accounted for. Specifically, evaporation processes are described by the algorithms of Stiver and Mackay \cite{stiver1984evaporation}, whereas dispersion and emulsification processes are represented using the algorithms by Mackay et 
al. \cite{mackay1980oil}. Finally, the influence of surface winds on the motion and deformation of the oil slick was incorporated using a wind coefficient of $3\%$ \cite{le2012surface}.

\subsection*{Experimental Setup}

As briefly discussed below, the present study adapts the setup presented in \cite{mittal2021hazard, hammoud2023variance} to the region surrounding NEOM. The computational domain covers the northern Red Sea region, extending across the longitudes $32^\circ$ to $37^\circ$ and latitudes $25^\circ$ to $30^\circ$ and up to a depth of approximately 2746~m . 
The domain is discretized using a computational mesh that is uniform in horizontal planes and non-uniform in the vertical direction.  It uses 500 equally-spaced nodes along the longitudinal axis, 500 equally-spaced nodes along the latitudinal axis, and 50 layers in the vertical direction.  The horizontal grid resolution is approximately 1~km.

From the met-ocean fields outlined above, the 3D ocean currents, surface winds, wave height and wave period from the years 2013 till 2017 were extracted and used an inputs to drive MOHID.  The Lagrangian particle transport model and weathering processes were solved using time steps of size 3600 s and 60 s, respectively.

\subsection*{Risk Quantification}
\label{sec:RiskMetrics}

The risks of individual oil spill sources are quantified in terms the arrival times of oil particles, and the volume fractions of oil beached on the NEOM shorelines. The arrival times represent the minimum traveling time of oil particles from each release source to the NEOM shorelines.  For each source, the volume fractions reflect the ratio of oil volume beached to the volume initially released. The arrival times are divided into four classes, namely $< 7$ days, 7-14 days, 14-30 days, and $>30$ days (as surrogate for no arrival during the simulation period).  Similarly, the volume beached are divided into four classes, namely $>50 \%$ of the initial release, $25-50 \%$, $<25 \%$, and $0 \%$ (when no oil is beached).  The results are illustrated using pie charts that depict the frequencies of the classes considered.  When generated for individual months of the year, the charts represent the outcome of fifteen experiments, as three simulations per month are performed for the five consecutive years investigated.

A finer analysis is also conducted where, instead of considering the entire NEOM coastline, smaller segments (approximately 25-km wide) are considered around specific sites, namely The Line, Duba, Sharma, Gayal and Magna. For each site, a histogram of the volume fraction is generated showing, for each source and release event considered, the amount of volume beached classified (using colors) in terms of the corresponding arrival time class. The histograms provide key information regarding the severity of the potential release event, and the time frame within which mitigation actions need to be deployed to minimize the impact on coastal areas.

Finally, an aggregate probability of volume beached along a given shoreline ($p_{i}$) is computed as:
\begin{equation}
    p_i = \frac{\sum_{j=1}^{15} \mathcal{V}_{i, j}}{\sum_{k=1}^{37} \sum_{j=1}^{15}\mathcal{V}_{k, j}} ,
\end{equation}
where, $\mathcal{V}_{i,j}$ is the fraction of volume beached from release location $i$ for event $j$, such that the event $j$ is an enumeration on the release times. 
The aggregate probability of volume beached measures the contribution of a given release source with respect to all the release sources. 
This metric allows contrasting sources by ranking release source based on their likely impact on the NEOM shoreline.

% !TEX root = ./main_merged.tex

\section*{Results and Discussion}
\label{sec:results}
\subsection*{Risk analysis for the NEOM shoreline}
\begin{figure}[p]
    \centering
    \vspace{-1cm}
    \includegraphics[width=0.33\linewidth]{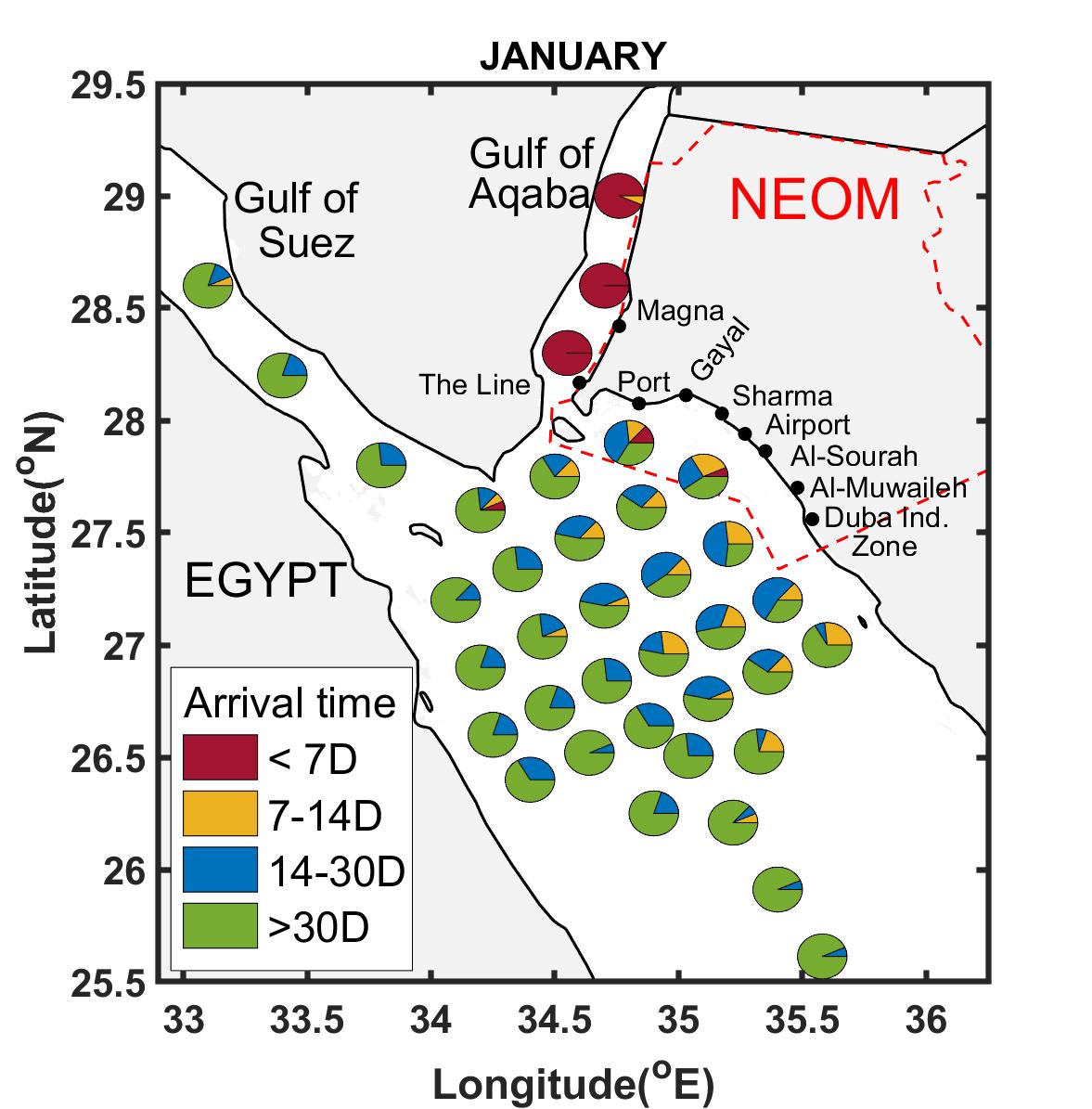}
        \includegraphics[width=0.33\linewidth]{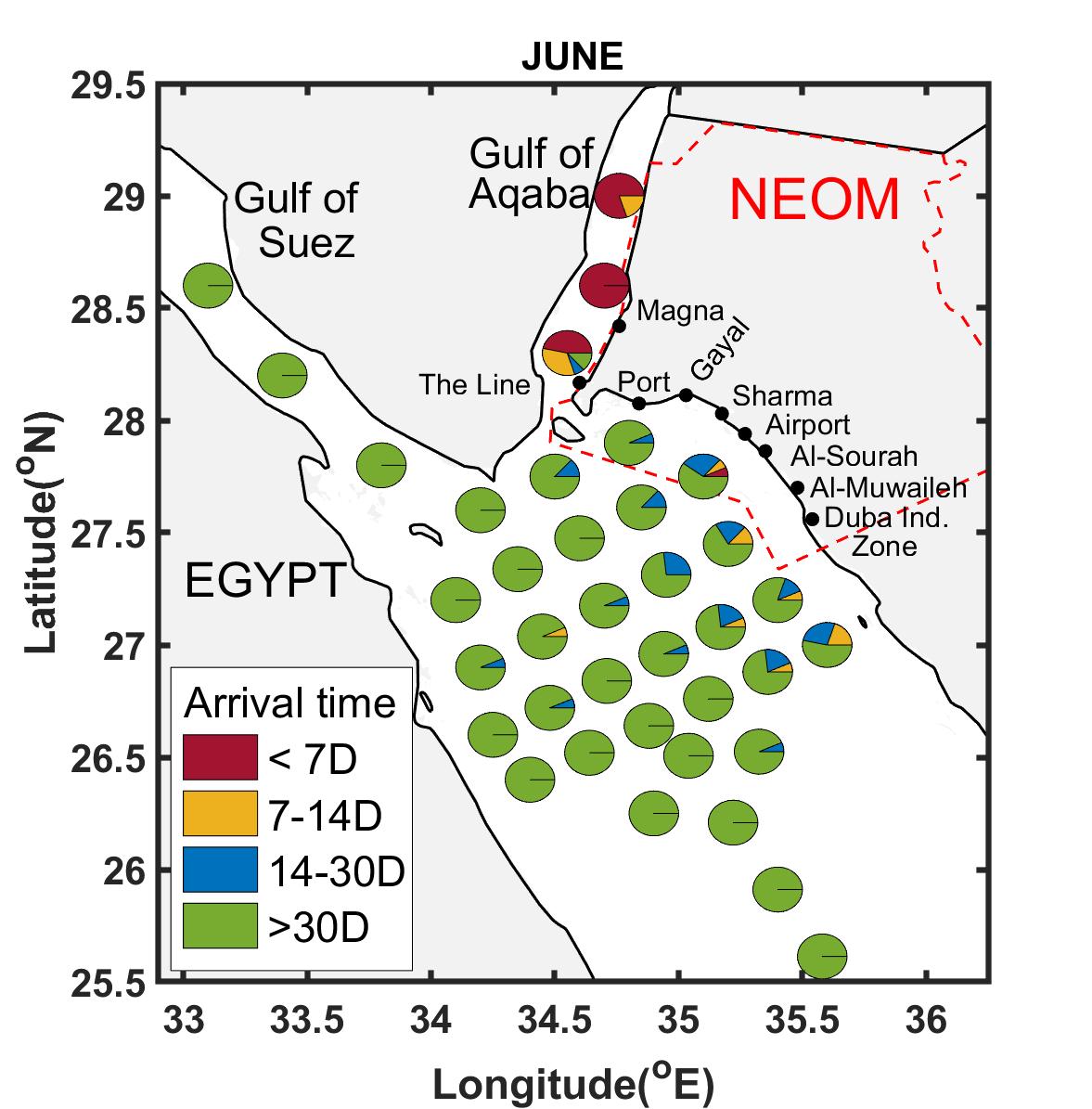}
            \includegraphics[width=0.33\linewidth]{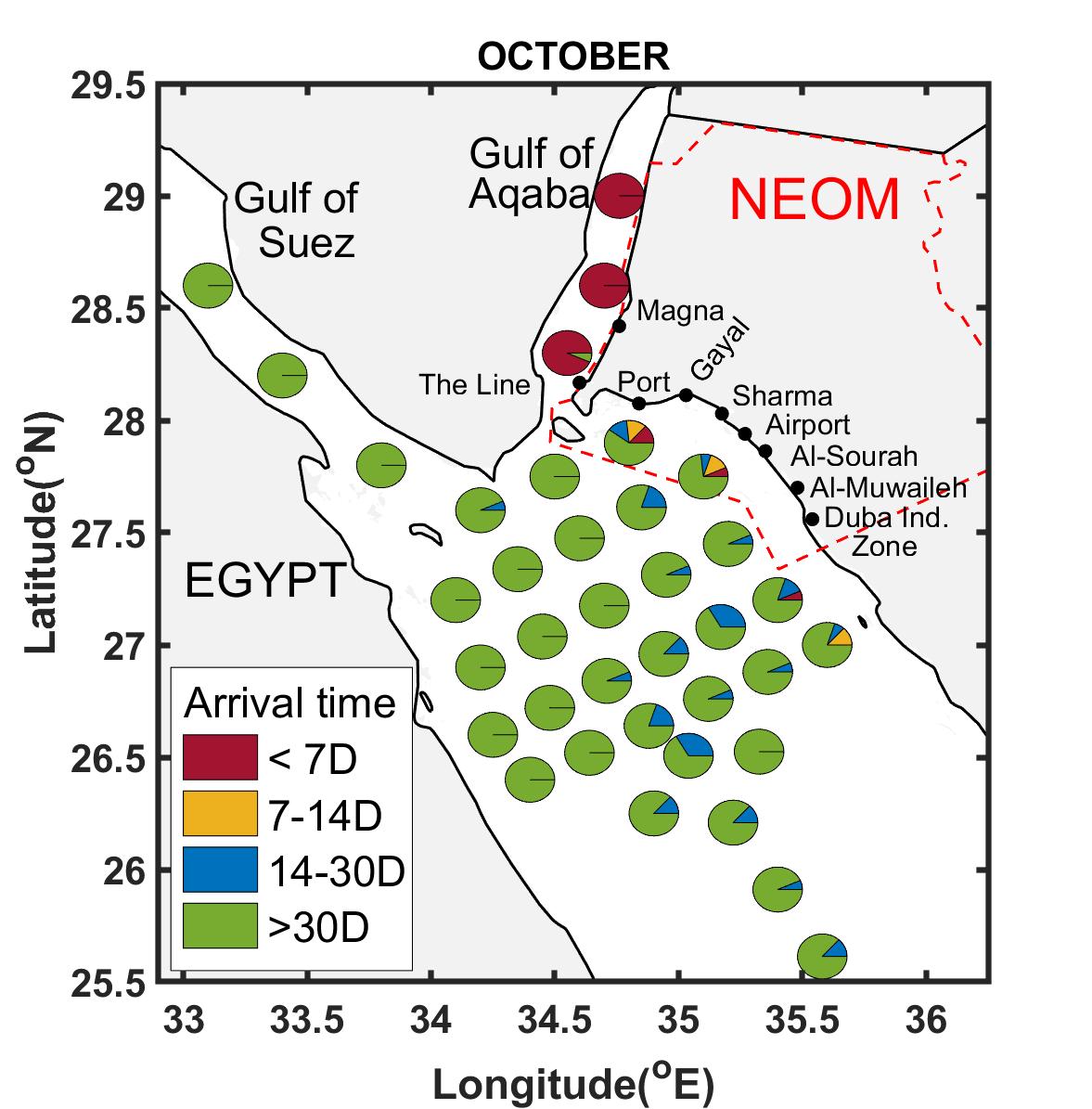}
             \includegraphics[width=0.33\linewidth]{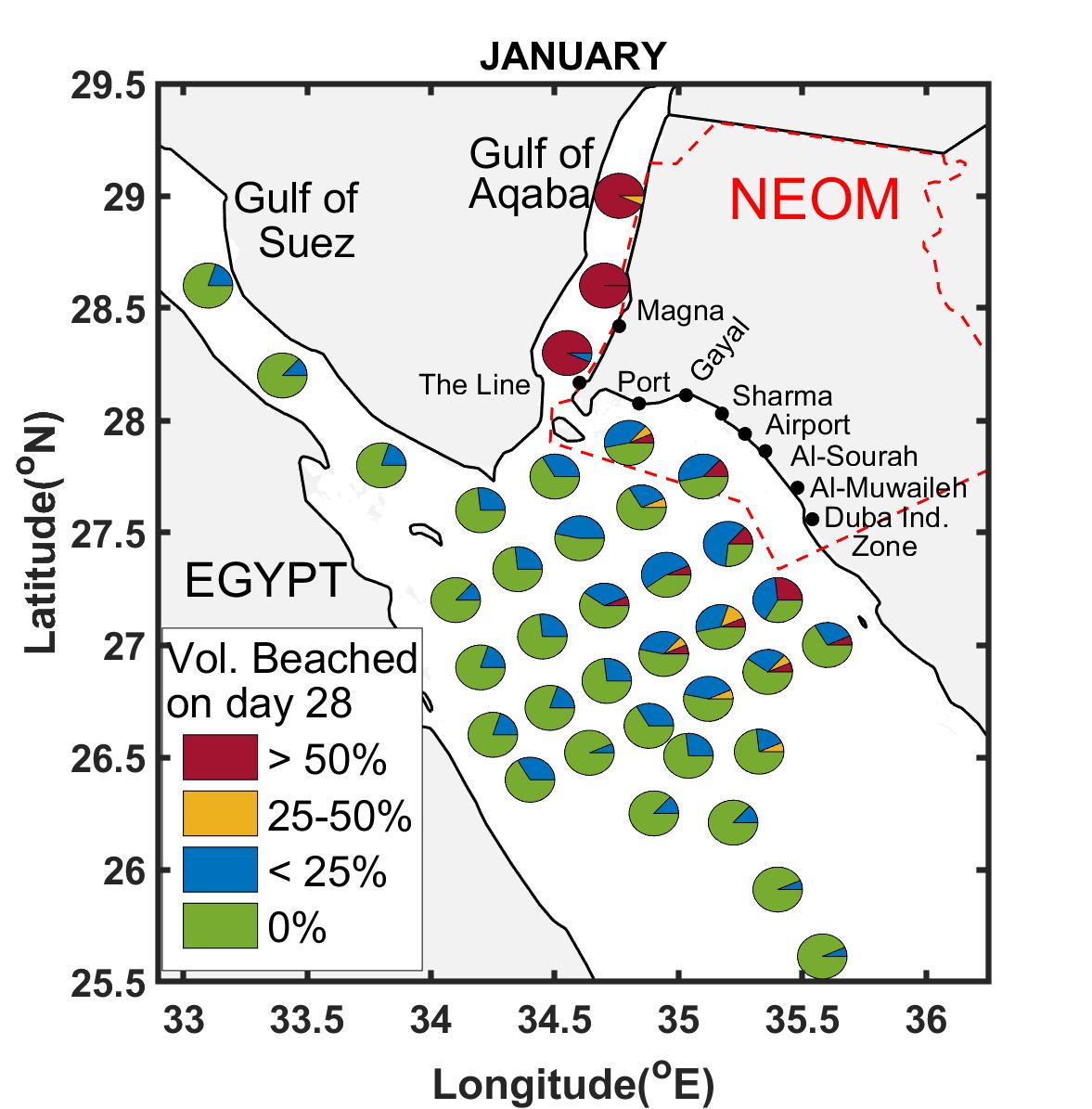}
        \includegraphics[width=0.33\linewidth]{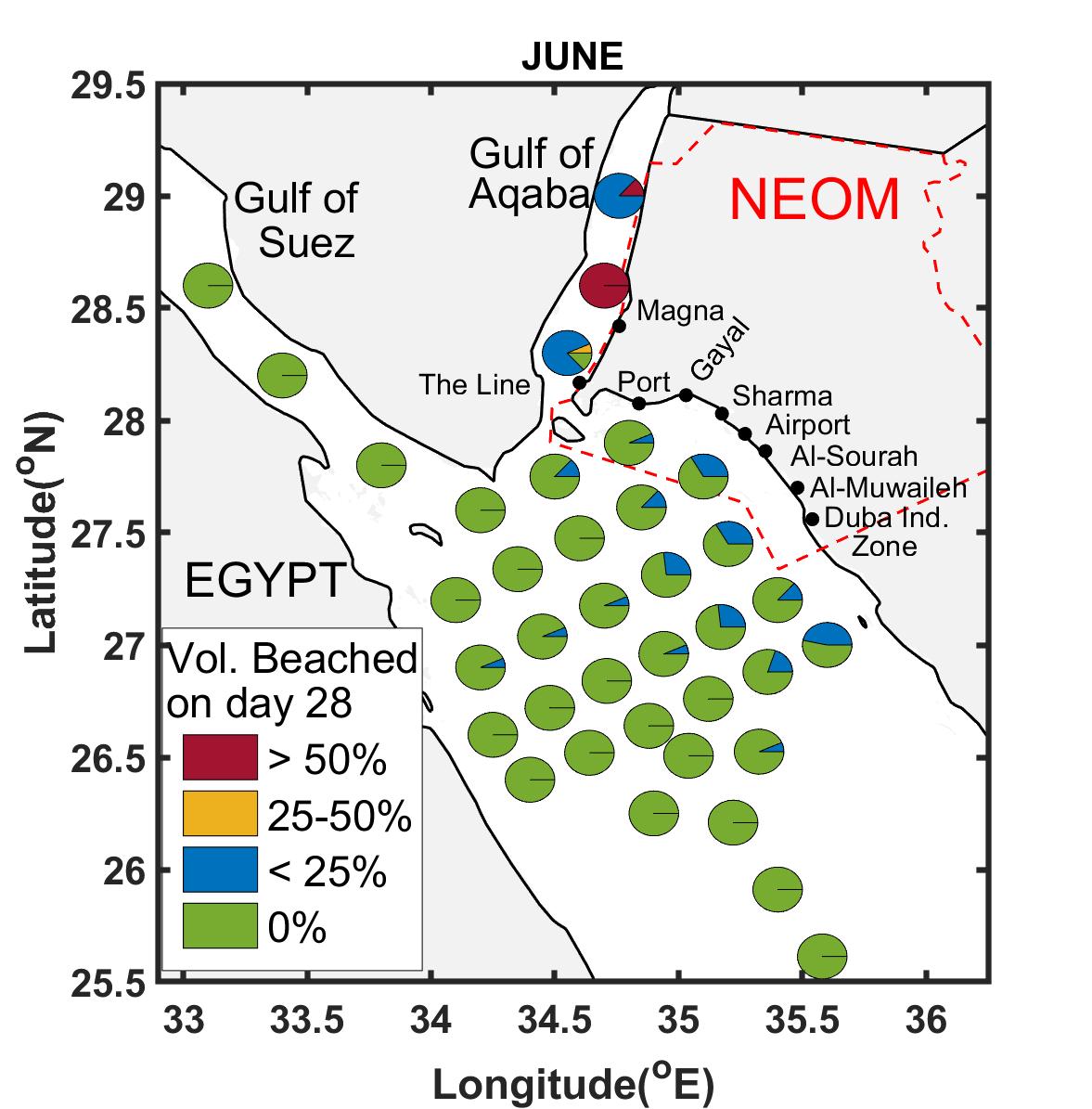}
            \includegraphics[width=0.33\linewidth]{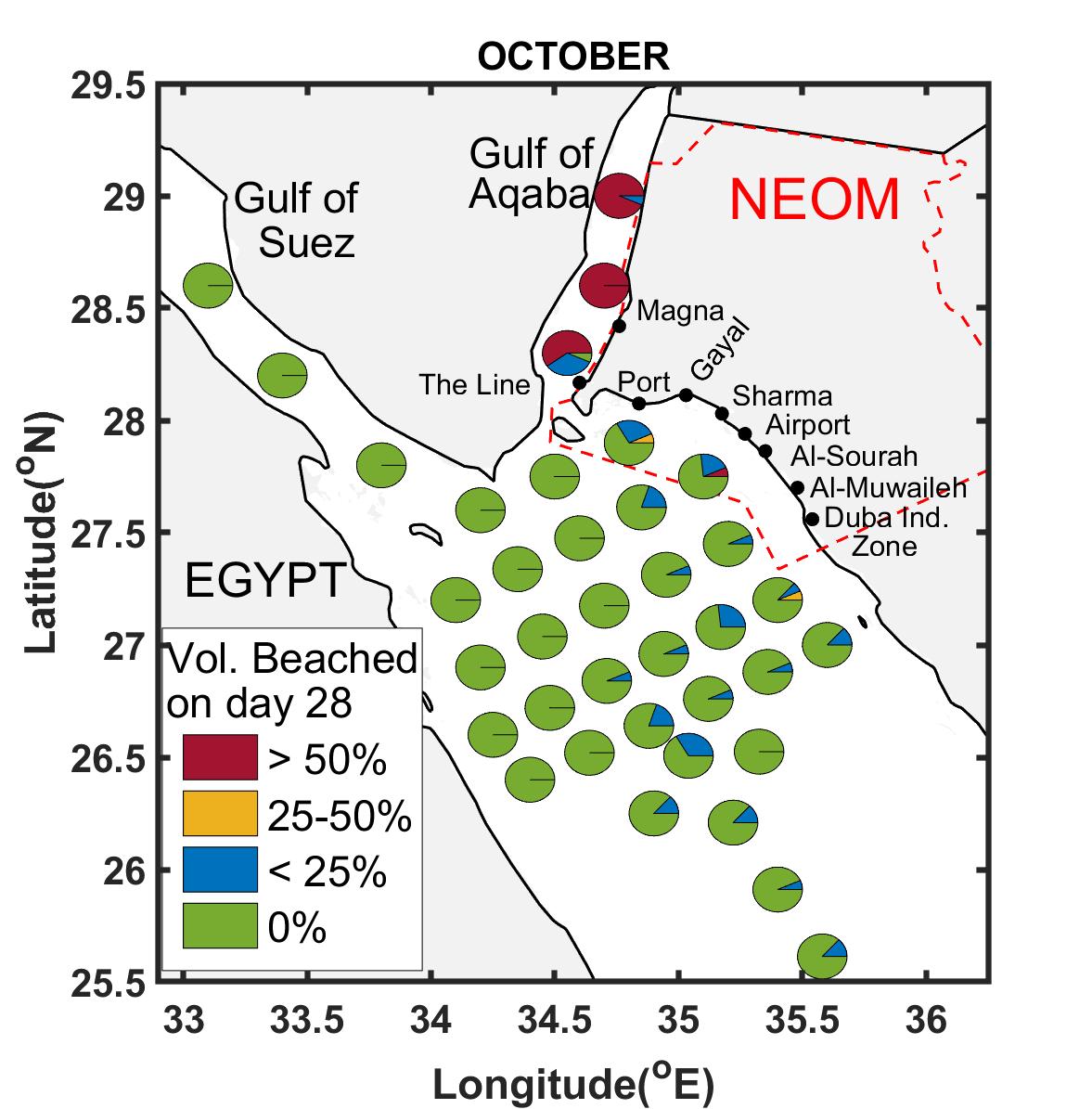}
\caption{Pie charts centered at each release source, representing the corresponding travel time of oil particles to the Neom shoreline (top row) and the volume fractions of beached oil particles (bottow row).}
    \label{arriv_vol_Neom}
    \end{figure}

\begin{figure}[t]
    \centering
    \includegraphics[width=0.33\linewidth]{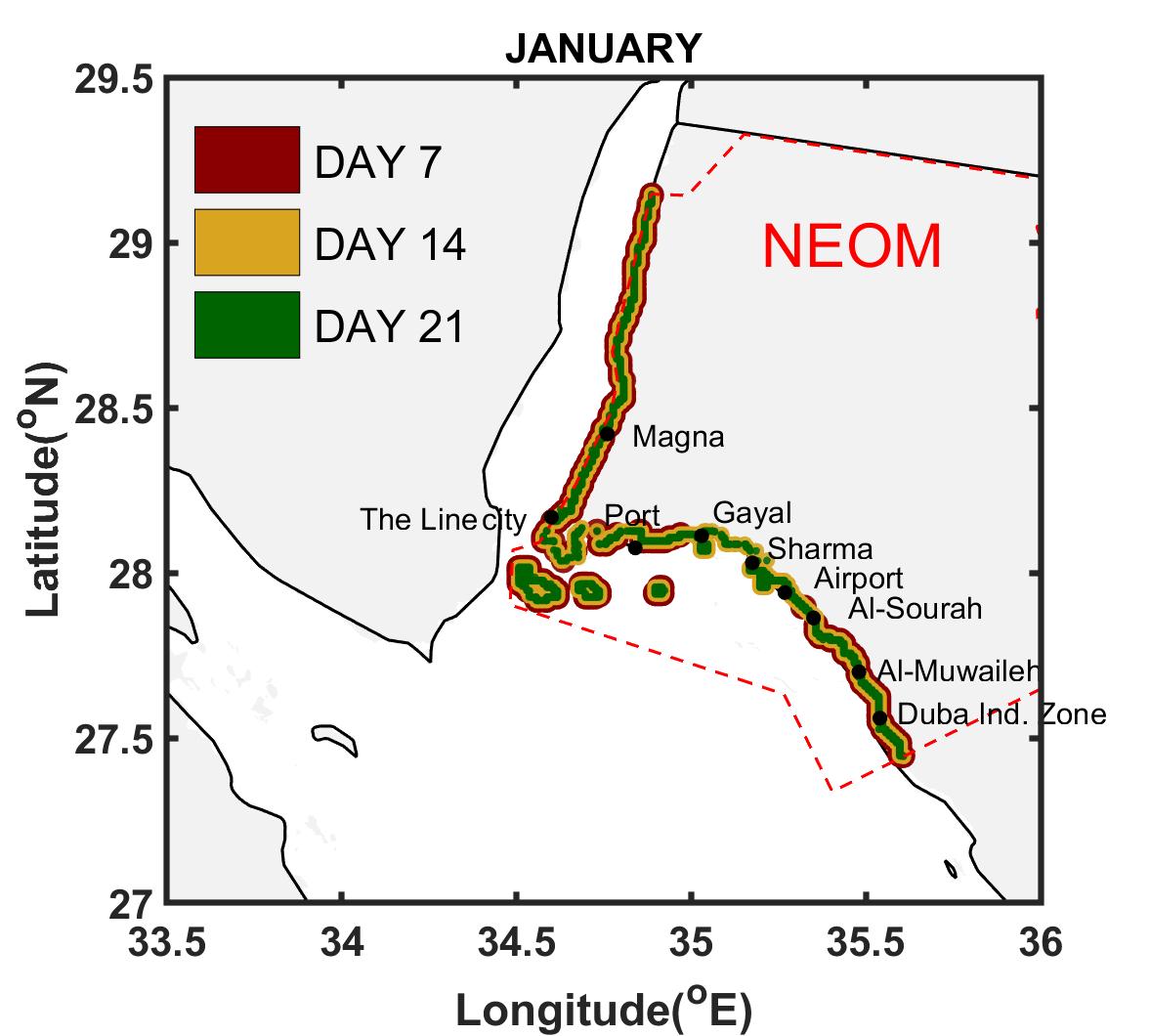}
    \includegraphics[width=0.33\linewidth]{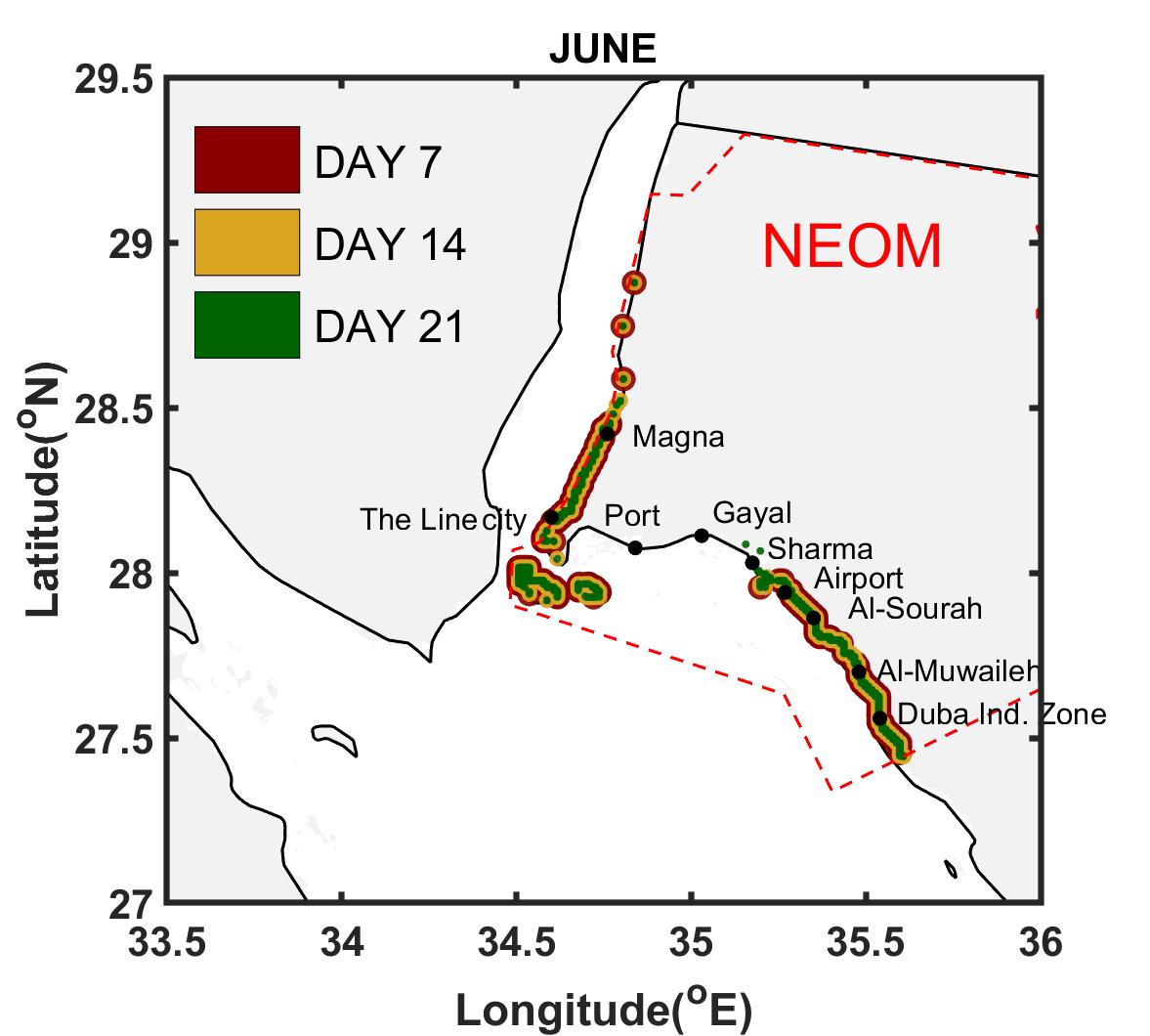}
    \includegraphics[width=0.33\linewidth]{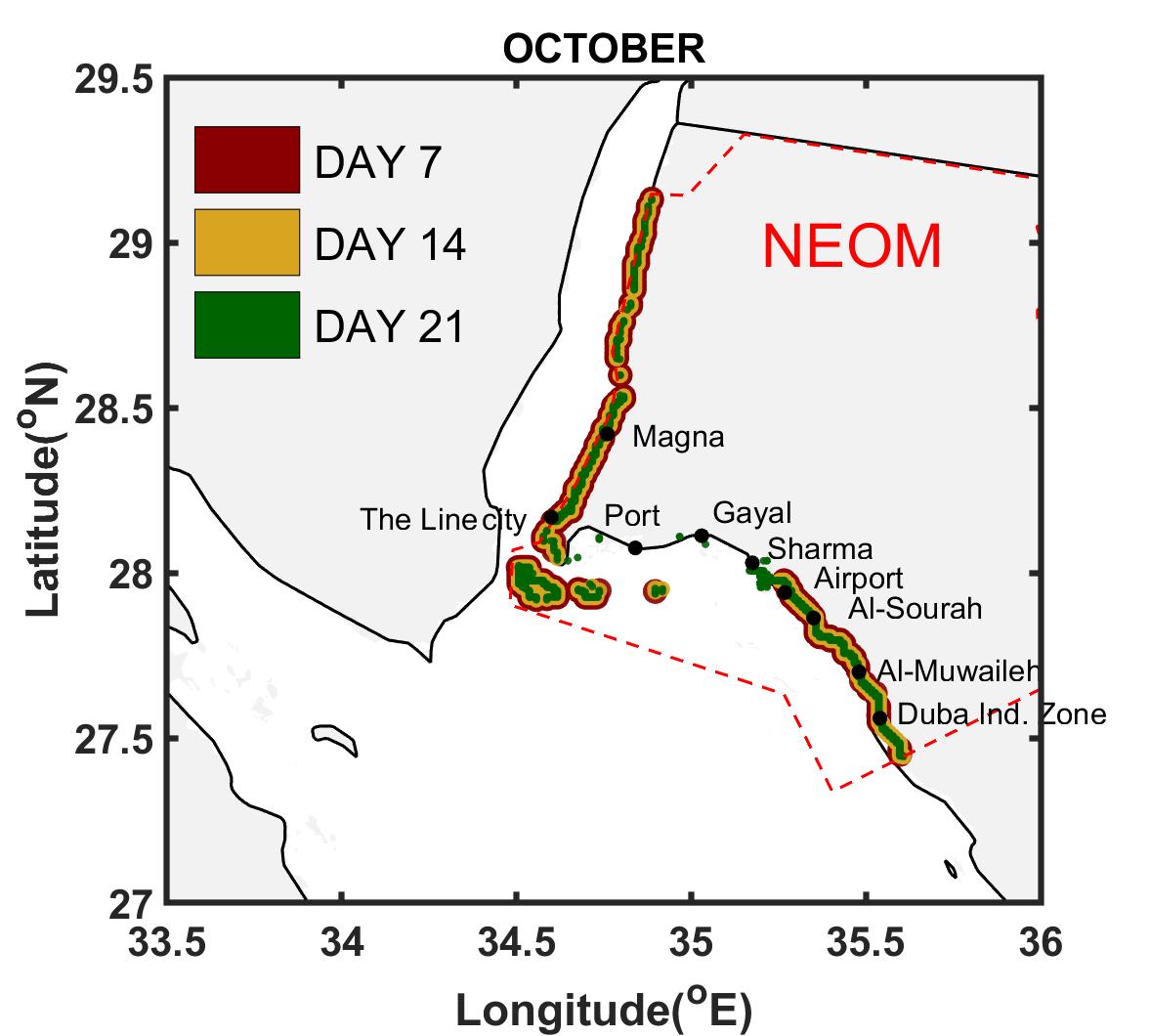}
    \caption{Regions of the Neom shoreline affected by beaching, for 7, 14 and 21 days after the onset of the spill.  Particles originating from all release sources are used to generate the contours. Plots are generated for release events occurring during the January, June and October months, as indicated.}
    \label{beached_all_main}
    \end{figure}

Figure~\ref{arriv_vol_Neom} and Supplementary Figures S1--S2 illustrate pie charts representing the impact of fifteen release events occurring during the months from January to December. The pie charts depict, for each release source, the travel time needed by the oil particles to reach the NEOM shoreline as well as the volume fraction of oil beached at the end of the simulation period.  Figure~\ref{beached_all_main} and Supplementary Figure S3 depict the region of the NEOM shoreline affected by beached oil particles, 7, 14 and 21 days following the release.  Particles originating from all release sources are used to generate these contours, thus illustrating the aggregate risk.  Release events originating during the months of January, June, and October are used for this purpose. 

Figure~\ref{arriv_vol_Neom} and Supplementary Figure S1 indicate that spills originating from sources $S_{35}-S_{37}$, which are located in the narrow Gulf of Aqaba and thus close to the shorelines, are characterized by short arrival times.  Within one week from the onset of the spill, entire segments of NEOM shoreline adjoining the Gulf of Aqaba are generally impacted; this occurs for all scenarios except for a few releases occurring during the summer months. In the summer months, the prevailing southwards currents in the Gulf of Aqaba tend to push the oil slicks towards the Tiran and Sanfir islands. Therefore, some segments of shorelines, located north of Magna city, may be shielded.  Within one week from the time of the spill, over 50\% of the volume of oil released by sources ($S_{35}-S_{37}$) may generally beach on the NEOM shore.  This occurs over the whole year except for the month of June. In June, the volume fraction of oil released from source $S_{35}$ that beaches on the NEOM shore is less than $25 \%$ by the end of the first week, but may rise to around $50 \%$ by the end of the third week following the onset of the spill.

The arrival times of oil particles originating from most of the sources in $S_{4}-S_{8}$ are less than one week during the whole year except during the months of June (except $S_{7}$), September and October. The volume fractions of oil beached originating from  sources $S_{4}-S_{8}$ are less than $25 \%$ by the end of the first week, but may rise to greater than $50 \%$ within two weeks after the onset of the spill, during Jan-May, July and August. These volume fractions are seen to exceed 50\% by the end of the first week of the onset during the months of November and December. This transport of spilled oil towards the NEOM shorelines is promoted by a cyclonic eddy that dominates the circulation in the Northern Red Sea region during the winter seasons.\cite{mittal2021hazard}

For the majority of release sources $S_{19}-S_{29}$, located in the open waters and close to the Egyptian coast, the arrival times fall in the interval of two to three weeks from the onset, for the months of November-March and July.  By the end of third week after the onset of the spill, the volume fractions of oil  originating from these sources remain below $25 \%$.  During the remaining months, only a few of the sources $S_{19}-S_{29}$ could impact the NEOM shorelines. Furthermore, the volume fraction of oil beached is less than $25 \%$, with relatively longer arrival times of around four weeks or no beaching in some scenarios. 

For sources $S_{32}-S_{34}$, which are located in the Gulf of Suez, a measurable impact on the NEOM shoreline is only observed during the months of  January-May and July. Beaching of oil originating from $S_{32}$ is recorded after week one during February, within one-two weeks in March and in May, two-three weeks in January and July.  Oil released from $S_{33}$ impacts the NEOM shorelines within two-four weeks in May and from January-March. For $S_{33}$, the arrival times fall within two-three weeks in January and three-four weeks in April and May.  The volume fraction of oil released by sources $S_{32}-S_{34}$ and beached on the NEOM shore remains less than 25\% by the end of the fourth week, following the onset of the spill. 

Figure~\ref{beached_all_main} shows that the NEOM shoreline extending from The Line in the north to Duba in the South is impacted in its entirety during January to May, but during June to December some segments are not significantly impacted. Specifically, by the end of the third week after the onset of the spill, beaching on the shoreline between Sharma and The Line is not predicted during June and from September to October.  Additionally, beaching on the shoreline between The Port of NEOM and The Line is not observed from May to September. The energetic meso- and submeso-scale circulations and northwesterly winds in the northern Red Sea region tend to split the oil slicks into different fragments.\cite{mittal2021hazard} These fragments are then transported in the opposite directions, towards both the Egyptian and Saudi Arabian shorelines, thereby sparing some segments between The Line and Sharma from beached oil during the months from June-December. 

\begin{figure}[b]
    \centering
    \includegraphics[width=0.33\linewidth]{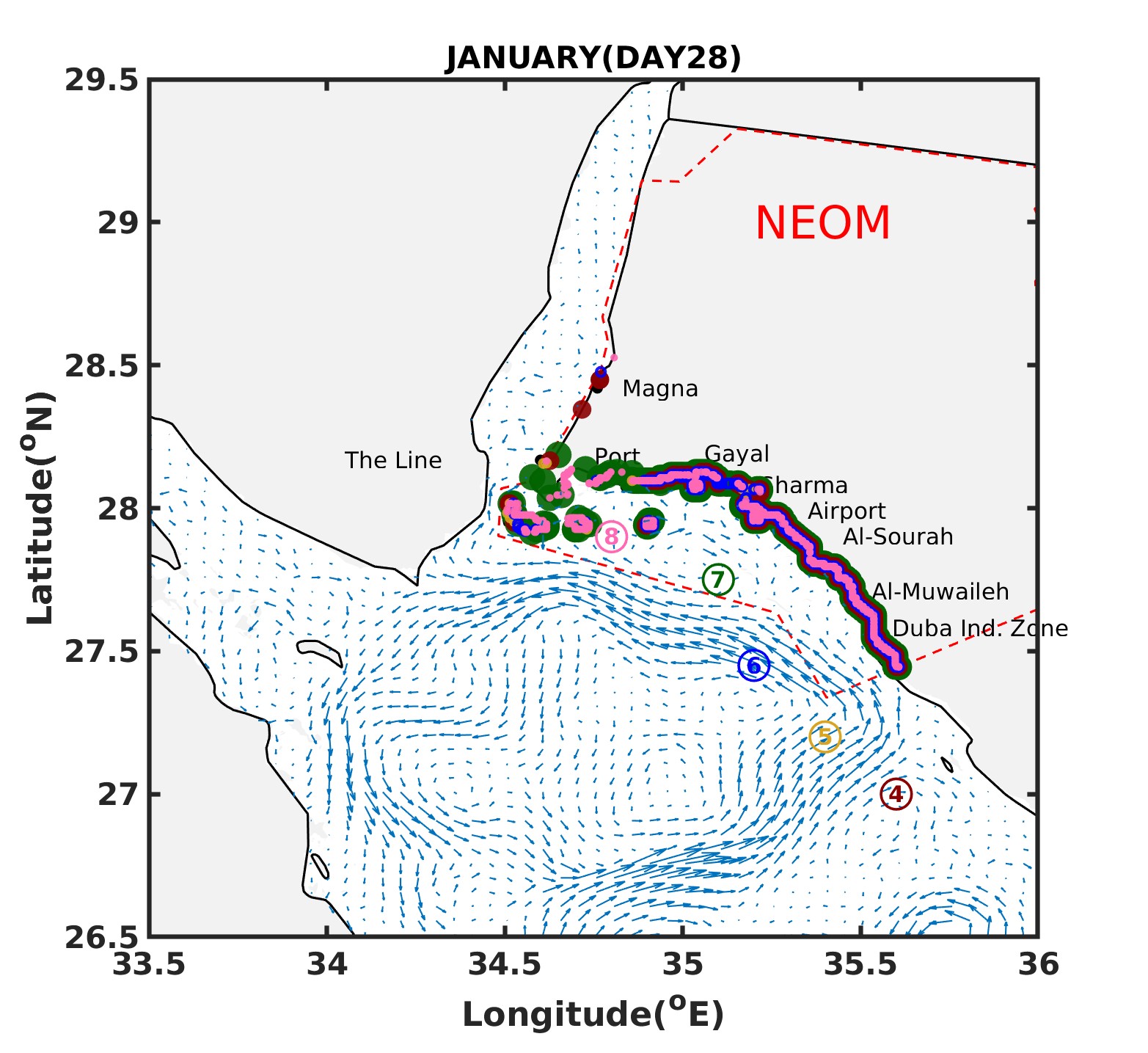}
    \includegraphics[width=0.33\linewidth]{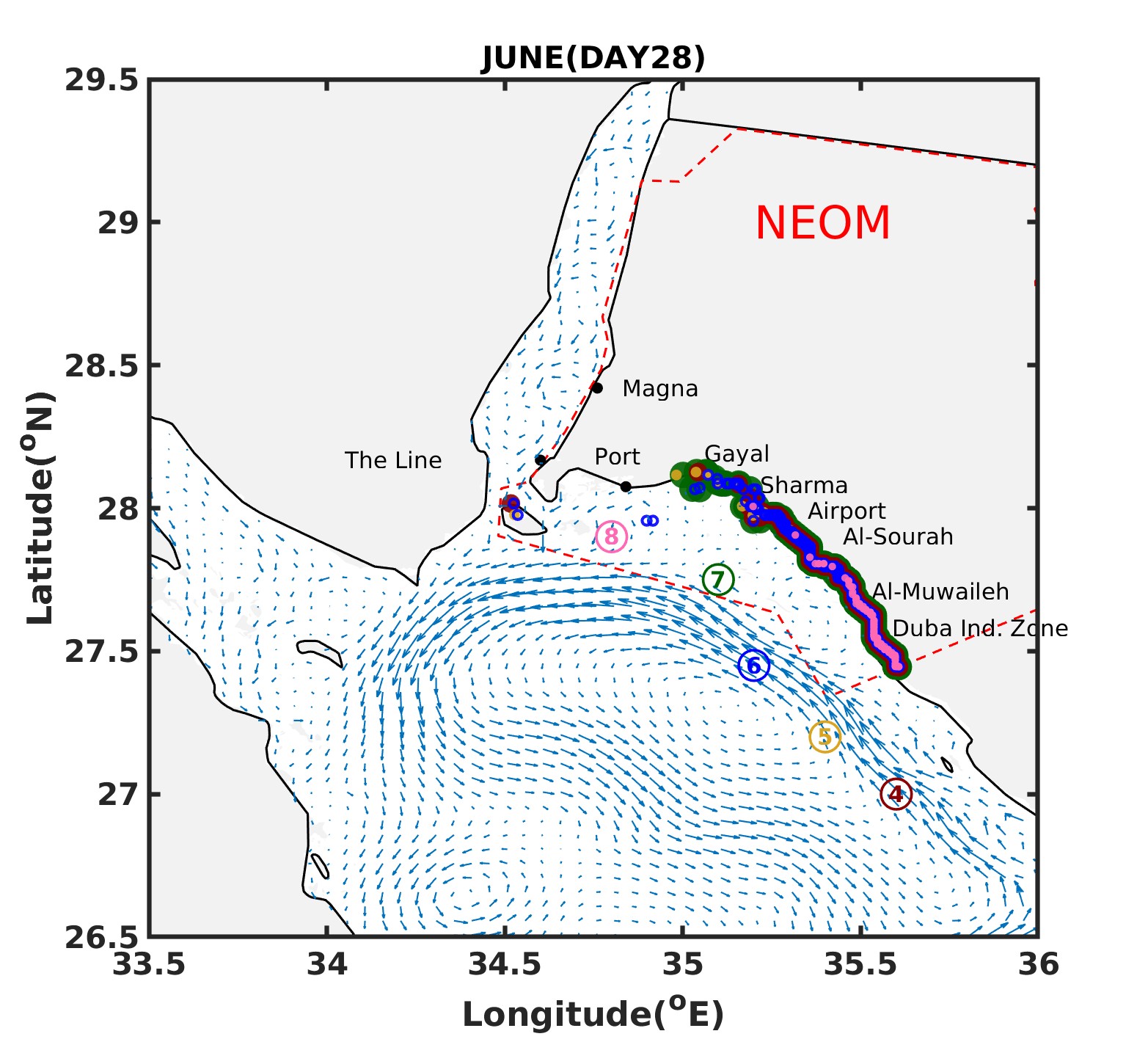}
    \includegraphics[width=0.33\linewidth]{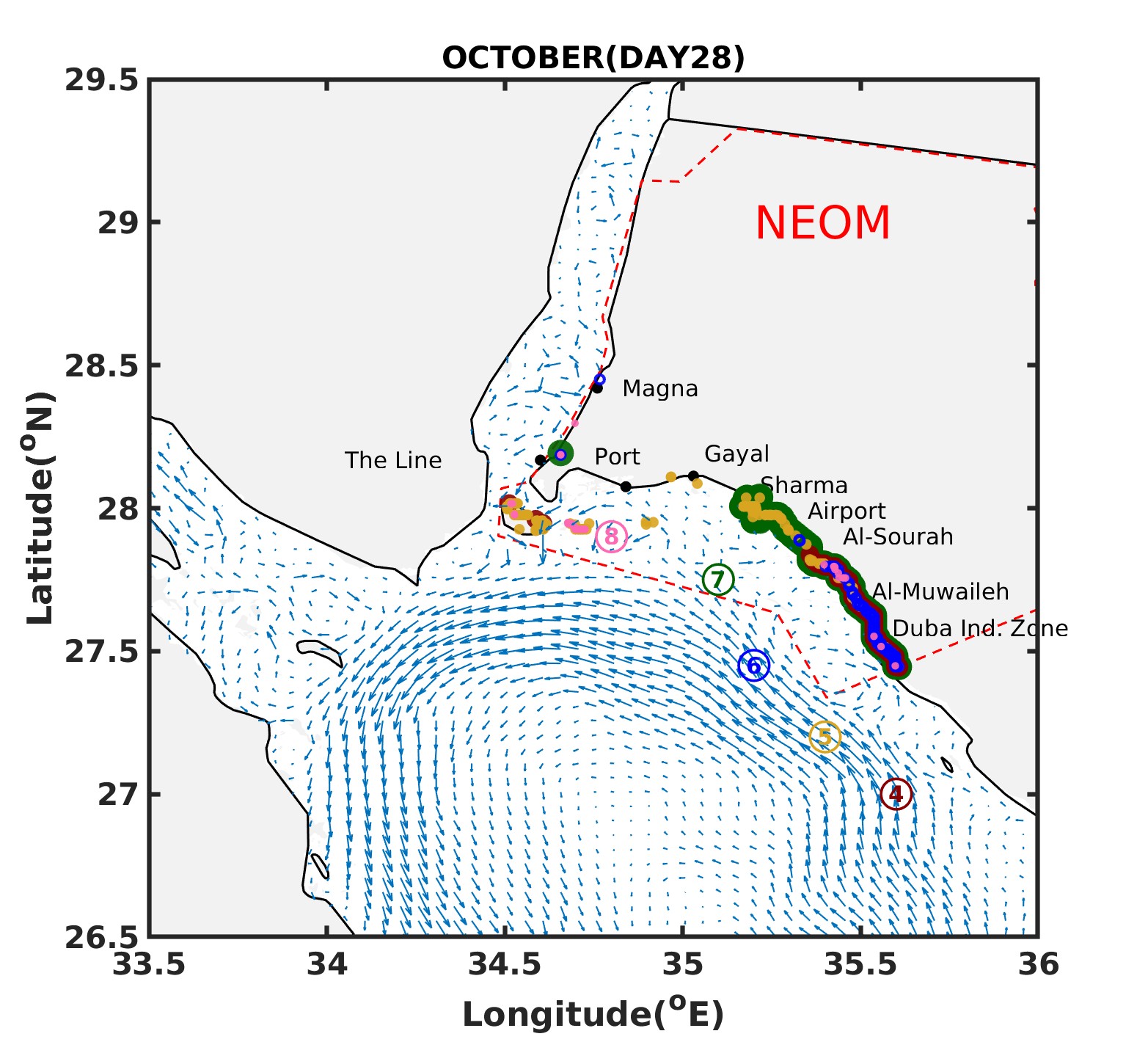}
    \caption{Regions of the Neom shoreline affected by beaching at the end of the 28-day simulation period. The contributions of selected sources
    are isolated using different color scheme for the individual sources, as indicated.  Plots are generated for release events occurring during the January, June and October months, as identified by the titles.}
    \label{beached_near_main}
    \end{figure}

Figure~\ref{beached_near_main} and Supplementary Figure 4 isolate the contributions of release sources $S_{4}-S_{8}$ which lie inside the NEOM boundary and are closest to its coastline. For these sources, beaching on the shorelines adjoining the Gulf of Aqaba is not observed in June and from August to October. For the remaining months, a measurable impact is observed on the shorelines adjoining the Gulf of Aqaba, from oil particles originating from $S_4$ (January-March and May), $S_{5}-S_{6}$ (February), $S_7$ (February-May, July and October-December) and $S_8$ (February, March and May). A substantial impact on the NEOM shoreline extending from The Line to Sharma is observed from the oil particles originating from $S_{4}$ (November), $S_5$ (October-November) and $S_7$ (October and December). Additionally, beaching of oil on the segment extending from the airport to Duba is not observed for $S_8$ during (January-May, November and December) and for $S_5$ during (January-August).  Overall, the results indicate that individual sources near the coastal may have severe impacts away from their location, as measured by the volume fraction of oil beached, and their impact may strongly depend on the seasonal variations of meto-cean conditions.

\subsection*{Risk analysis for specific sites}
The risk associated with the individual release sources is now analyzed for specific sites along the Neom coast, namely The Line, Duba, Sharma, Gayal and Magna. Figures~\ref{Line_histo},~\ref{Duba_histo} and Supplementary Figures S5--S9 plot the histograms of volume fractions for each source and release event considered, showing the amount of volume beached and the corresponding arrival time class (classified using colors), during the months from January to December. Figures~\ref{Line_prob},~\ref{Duba_prob} and Supplementary Figures S10--S12 depict the (inverse) risk probabilities for each of the specific sites considered. These probabilities characterize the region of dependence of the spill risk, as estimated using Eq. 1.

\subsection*{\textit{The Line}}
\begin{figure}[p]
    \centering
    \includegraphics[width=0.8\linewidth]{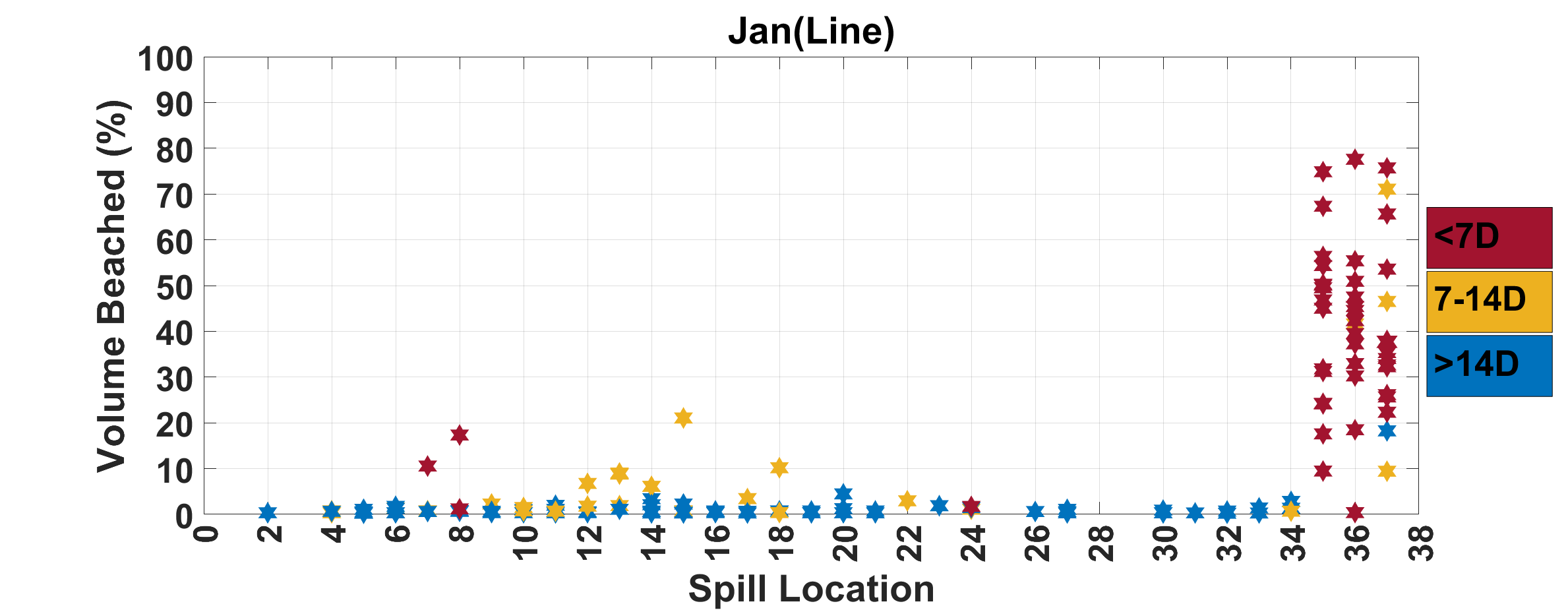}
    \includegraphics[width=0.8\linewidth]{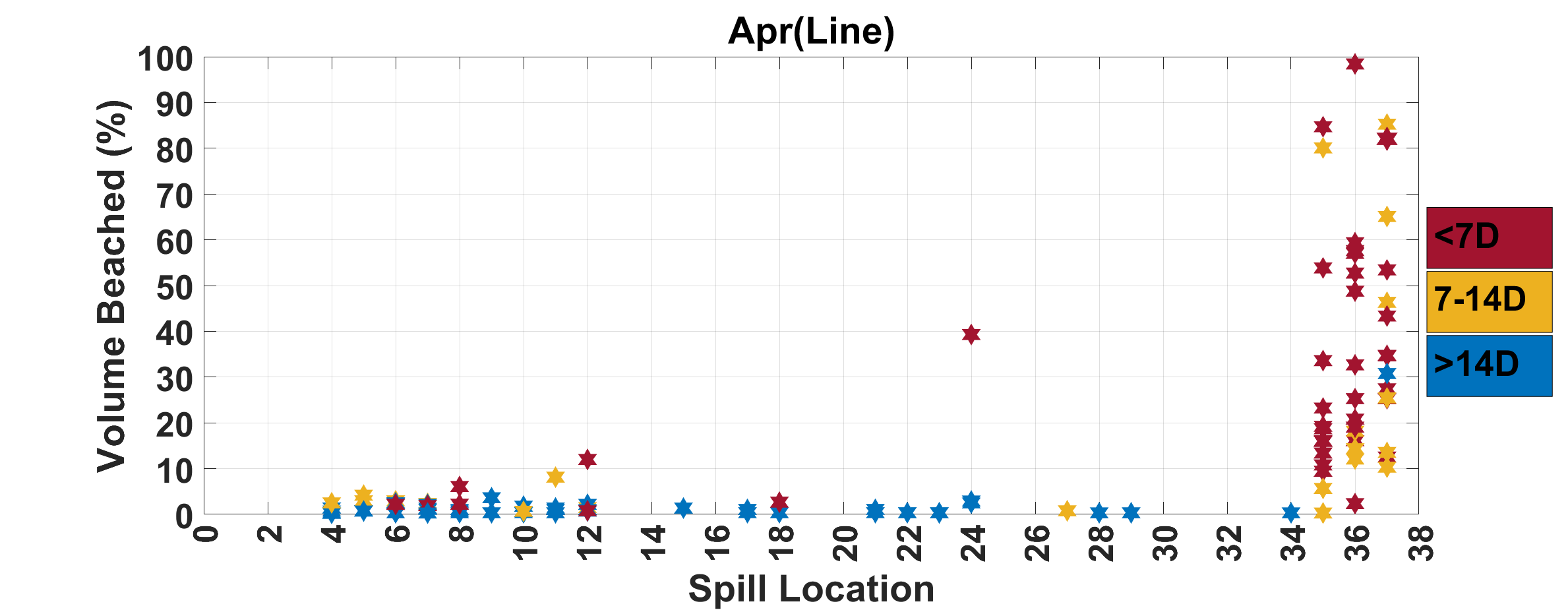}
    \includegraphics[width=0.8\linewidth]{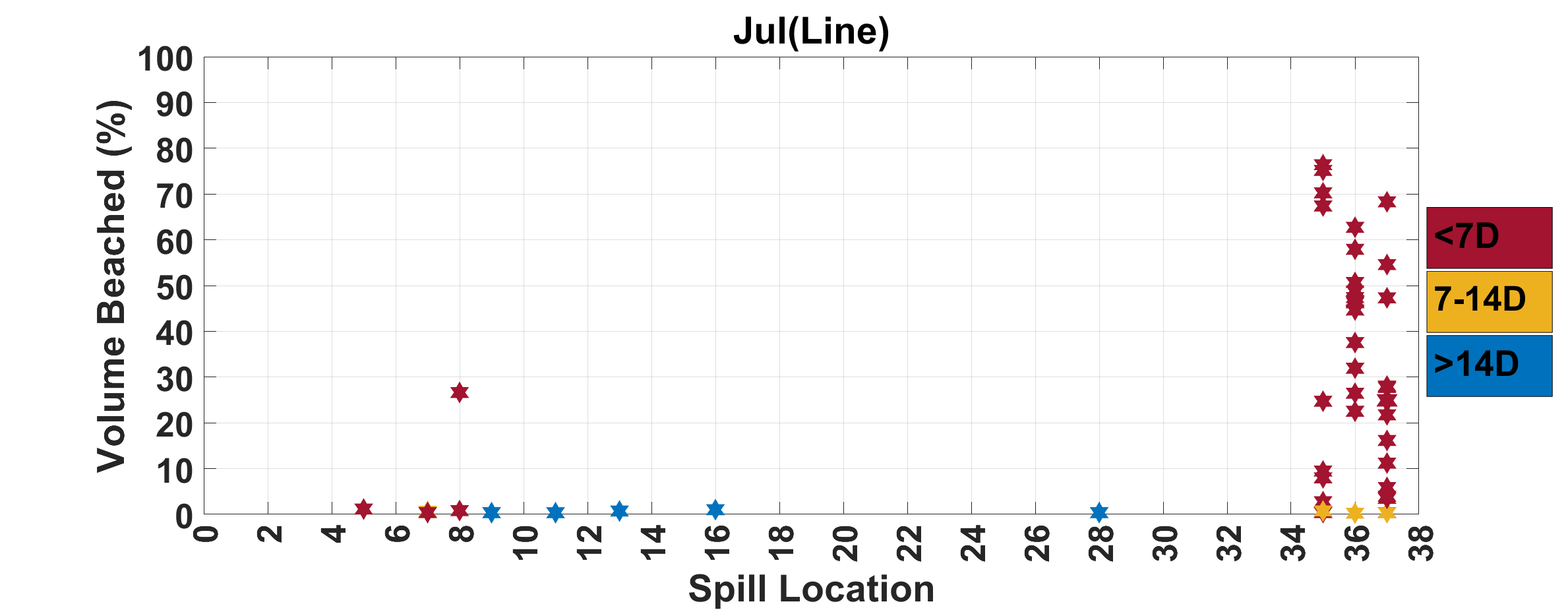}    \includegraphics[width=0.8\linewidth]{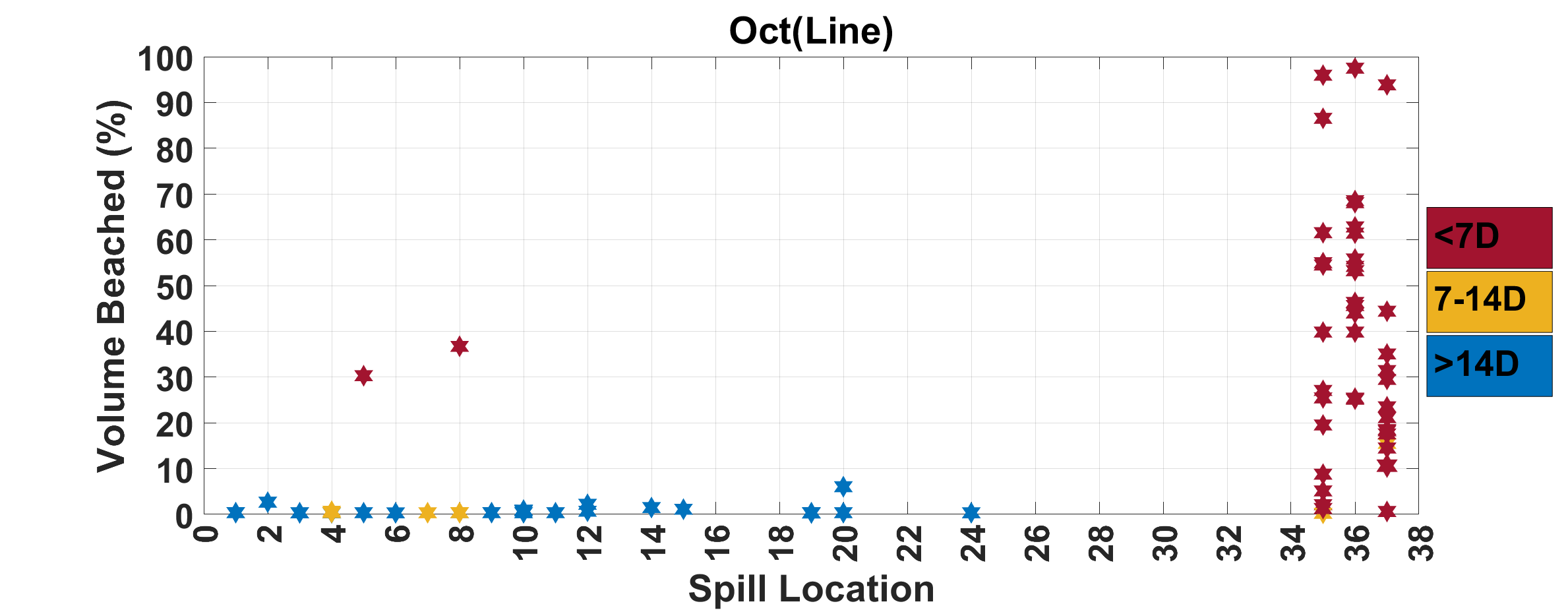}
    \caption{Histograms of the volume fractions beached at the shorelines of The Line.  Predictions from all release sources and events are
    classified (using colors) in terms of the corresponding arrival times. Plots are generated for release events occurring during the January, April, July and October months, as indicated.}
    \label{Line_histo}
    \end{figure}

Figure~\ref{Line_histo} and Supplementary Figure S5 plot histograms of the volume fractions beached at the shorelines of The Line, where predictions from all the release sources and events are classified in terms of the corresponding arrival times. The histograms present a uni-modal distribution of the volume fractions with tails varying from approximately 10\% to 80\%. The spills originating primarily from sources $S_{35}-S_{37}$ are characterized by the highest severity (low arrival times and high volume fractions) amongst other sources. During the months of April and from September-December, the volume fraction of oil released from source $S_{35}$ and beached around The Line may rise to 85\% by the end of the first week. The volume fraction of oil released from source $S_{36}$ and beached around The Line is greater than 60\% over the whole year except during April, June and October (greater than 90\%). The volume fraction of oil released from source $S_{37}$ is greater than 50\% throughout the year, except during the months of June and August (around 20\%), by the end of first week. The prevailing northwards currents\cite{mittal2021hazard} towards The Gulf of Aqaba tend to quickly push oil released from $S_8$ towards The Line; in March, the volume fraction may rise to more than 90\%. However, the volume fractions remain less than 50\% for the whole year except for March, June and September. The segments around The Line may be weakly affected by oil originating from $S_8$ in June and September. Additional events having early arrival times are associated with $S_{18}$ and $S_{24}$, which are located close to the northern tip of The Red Sea between The Gulf of Aqaba and The Gulf of Suez (near Sharm El-Sheikh). Here, the transport of spilled oil towards The Line is promoted by the prevailing coastal currents, which dominate the circulation during the months from December to May. The arrival times fall within one-two weeks during these months. Specifically, the arrival time is less than one week during December, February, and April for $S_{18}$,  and during January and April for $S_{24}$. Events with short arrival time (less than one week) are also associated with $S_{5}$ (in March and October) with volume fractions of around 40\%. However, very few sources among 
$S_9-S_{34}$ are characterized by moderate arrival times (two to three weeks), and generally have low severity in terms of amount of beached oil (volume fractions less than 10\%).

\begin{figure}[p]
    \centering
    \includegraphics[width=0.8\linewidth]{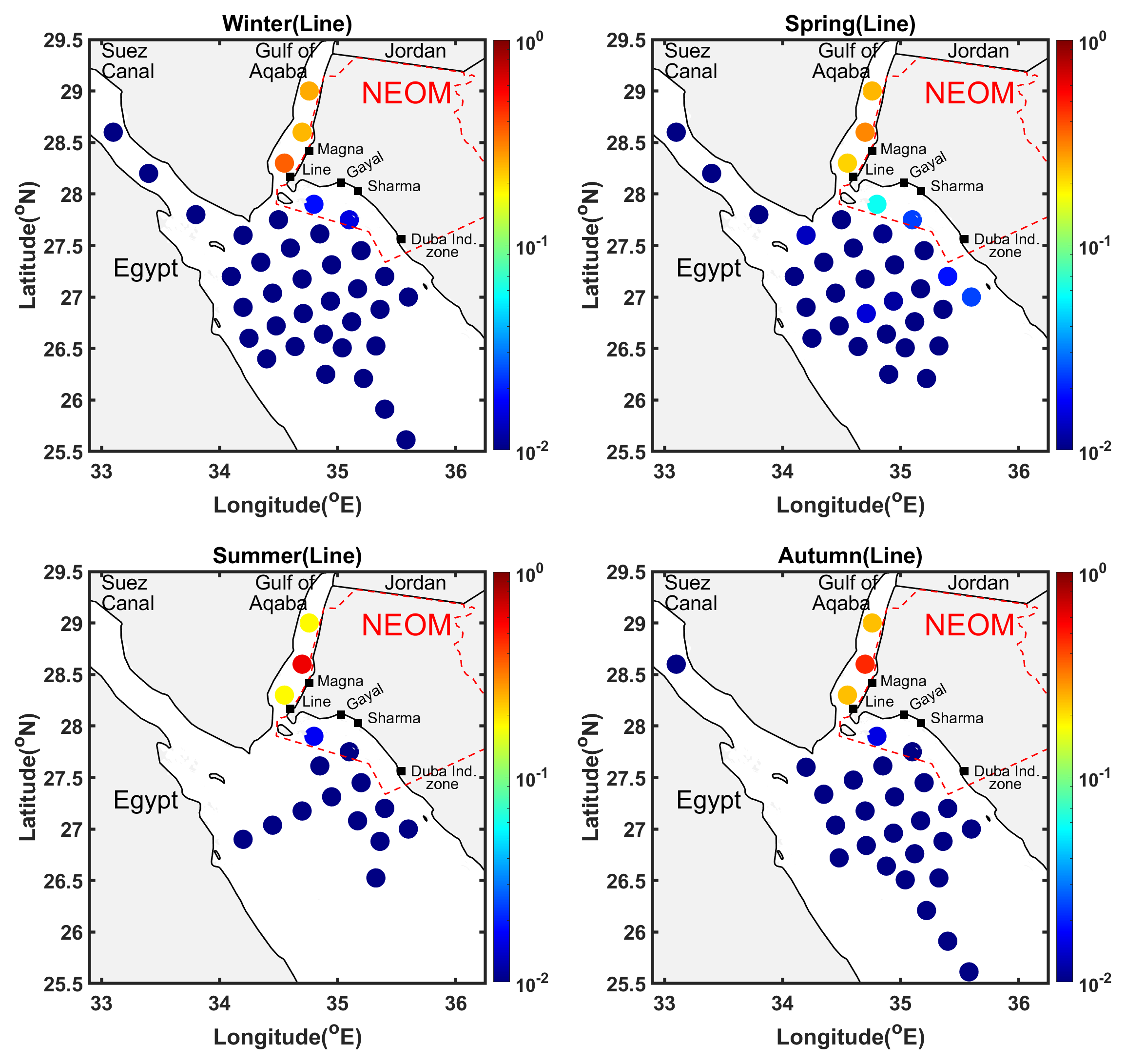}
    \caption{Risk probabilities for shorelines of The Line.  The probabilities, estimated using Eq.~1, characterize the region of dependence of the overall risk. Plots are generated for release events occurring during the four seasons, as indicated.}
    \label{Line_prob}
    \end{figure}

\begin{figure}[p]
    \centering
    \includegraphics[width=0.8\linewidth]{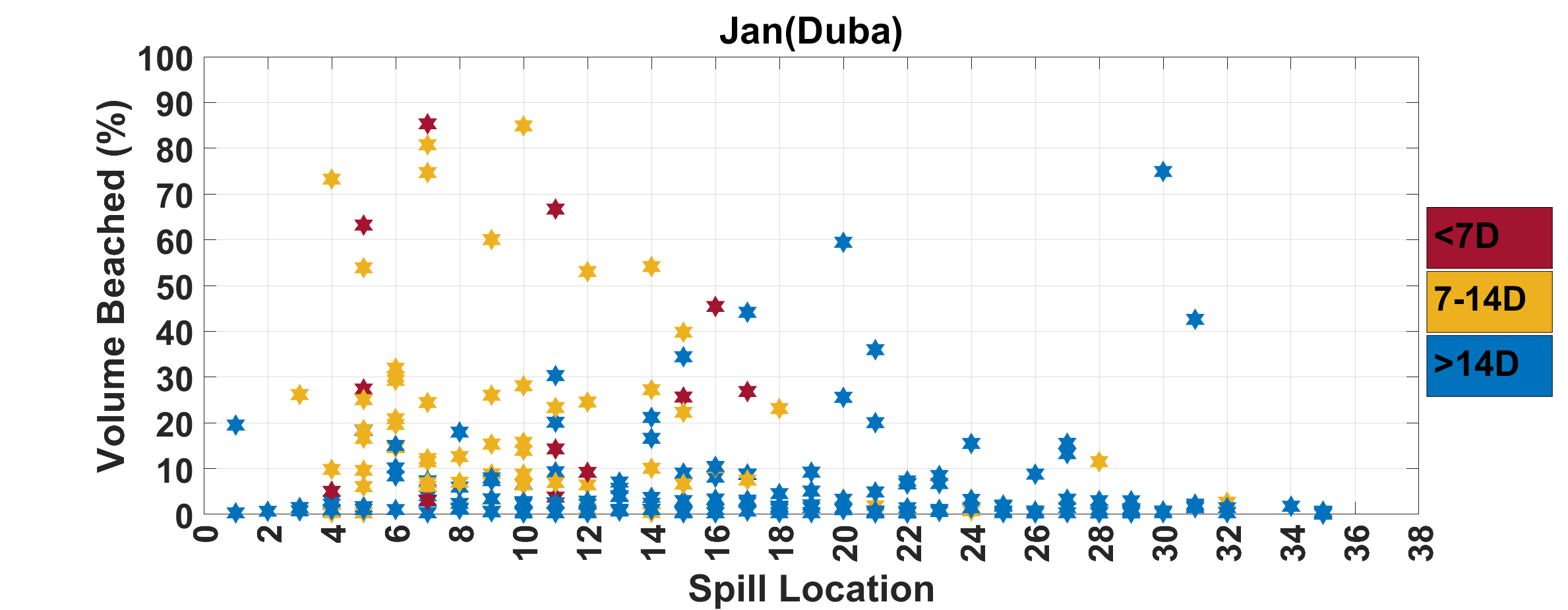}
    \includegraphics[width=0.8\linewidth]{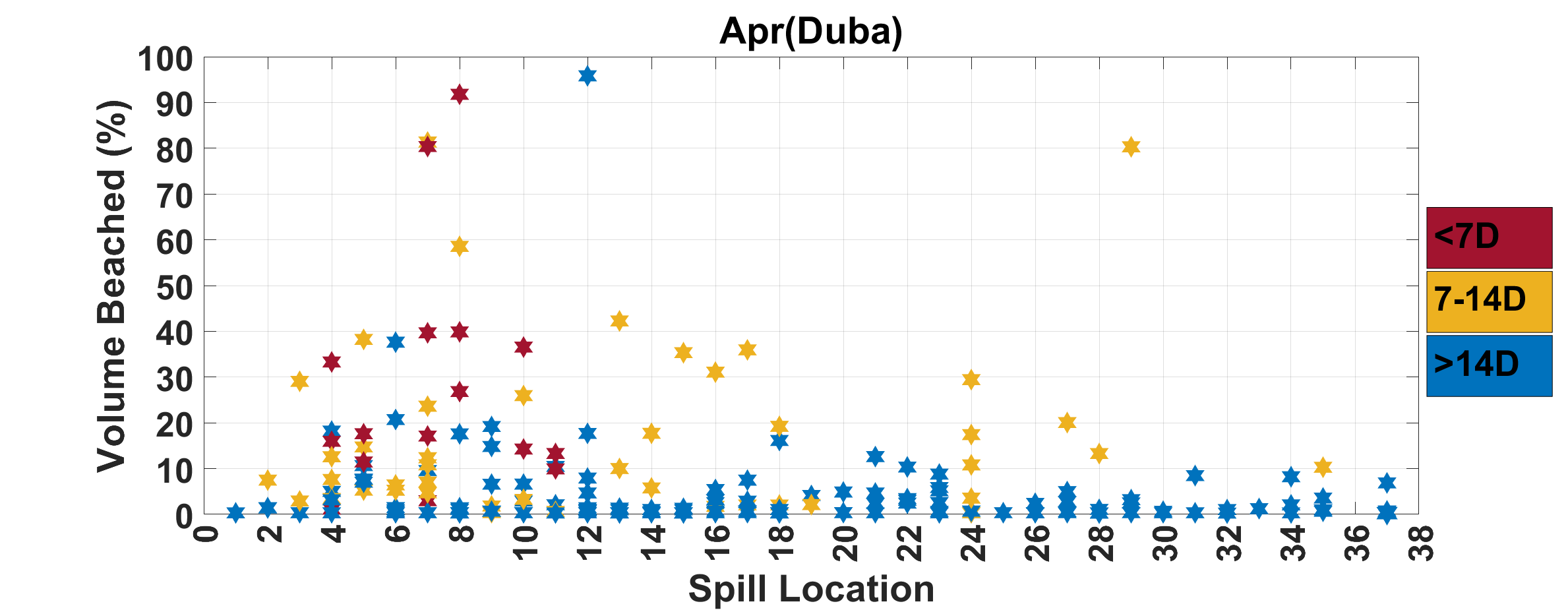}
    \includegraphics[width=0.8\linewidth]{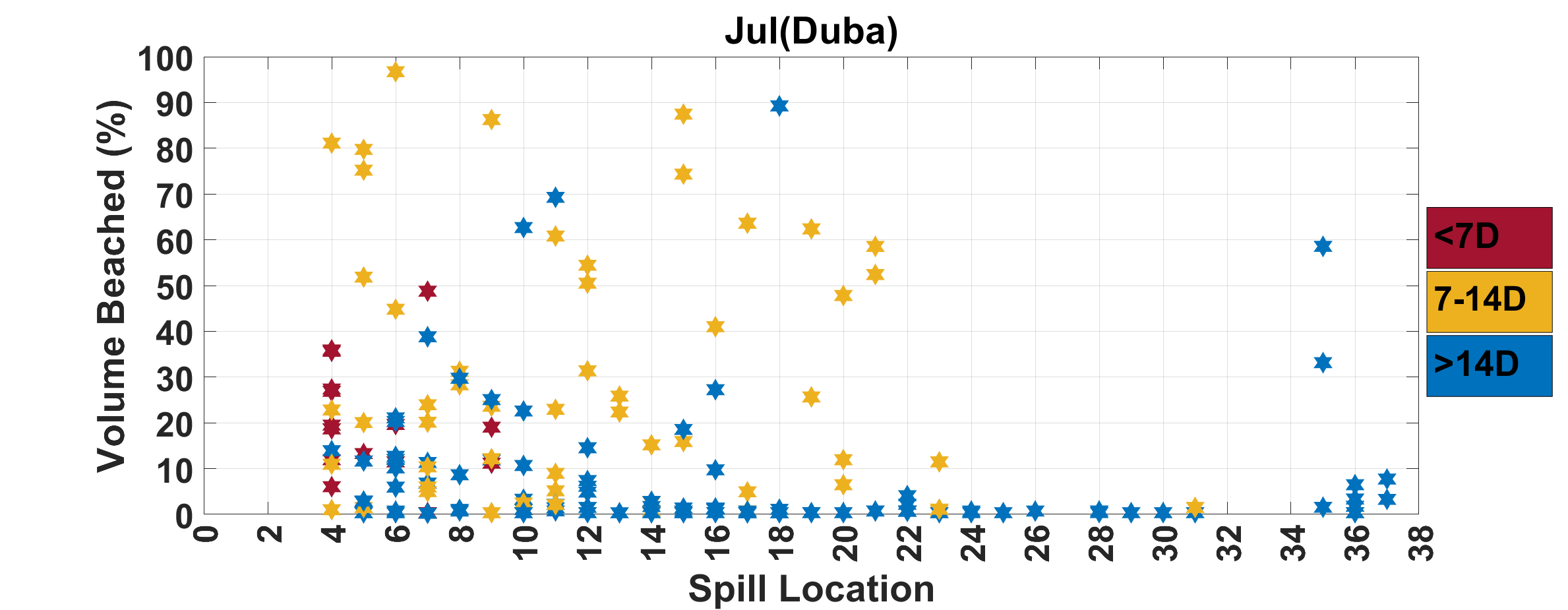}    \includegraphics[width=0.8\linewidth]{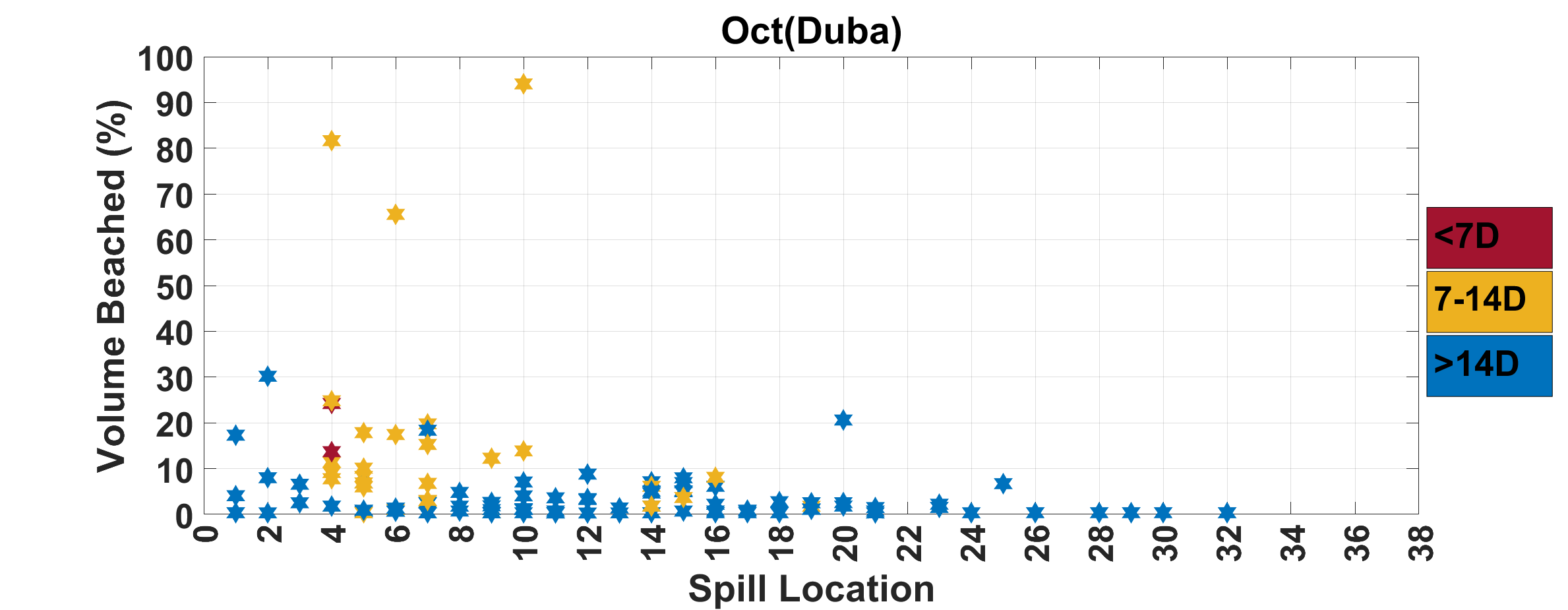}
    \caption{Histograms of the volume fractions beached at the shorelines of Duba.  Predictions from all release sources and events are classified (using colors) in terms of the corresponding arrival times. Plots are generated for release events occurring during the January, April, July and October months, as indicated.}
    \label{Duba_histo}
    \end{figure}

\begin{figure}[p]
    \centering
    \includegraphics[width=0.8\linewidth]{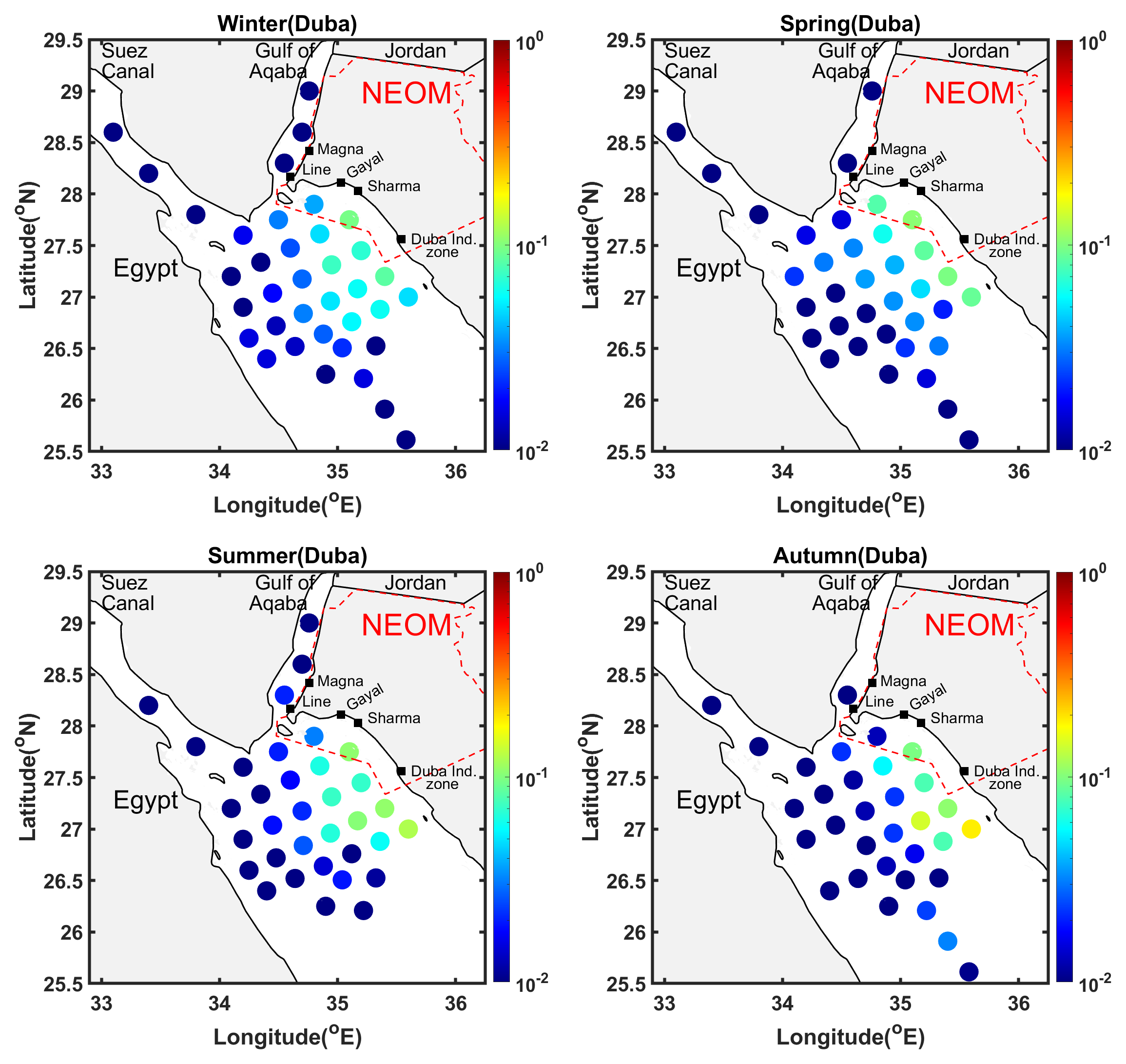}
    \caption{Risk probabilities for shorelines of Duba.  The probabilities, estimated using Eq.~1, characterize the region of dependence of the overall risk.  Plots are generated for release events occurring during the four seasons, as indicated.}
    \label{Duba_prob}
    \end{figure}

Figure~\ref{Line_prob} depicts the seasonal distribution of risk probabilities, estimated using Eq.~1 for oil beached around The Line. Sources $S_{35}-S_{37}$, located in The Gulf of Aqaba, are responsible for the largest risk.  
The risk associated with $S_{36}$ is the highest amongst $S_{35}-S_{37}$ in spring, summer and autumn seasons, whereas the risk associated with $S_{35}$ is highest in winter.  The risk associated with the remaining sources is appreciably smaller than that observed for $S_{35}-S_{37}$.
In addition, the associated probabilities are very small, except possibly for sources $S_5$--$S_8$ for which appreciable values may occur. 
Overall, the results of Figures~\ref{Line_histo} and~\ref{Line_prob} indicate that for The Line, the risk is primarily dominated by sources located in the Gulf of Aqaba, followed to a lower extent by sources located close to its shoreline.

Spills originating from sources located in the Gulf of Aqaba generally lead to severe events, with a large fraction of the oil released beaching within a short period (< 7 days) from the time of the release.  Consistent with the histograms in Figure~\ref{Line_histo}, sources located in the Red Sea and close to the NEOM shoreline may result in severe impact on The Line, but these events have low probability of occurrence, leading to small risk values reported in Figure~\ref{Line_prob}.

\subsection*{\textit{Duba}}

Figure~\ref{Duba_histo} and Supplementary Figure S6 plot histograms of the volume fractions beached at the shorelines of Duba.  In contrast to those corresponding to The Line, the results indicate that the shoreline surrounding Duba is vulnerable to sources located in the entire region
facing its coast.  This is reflected by the fact that multiple events with severe impacts are observed for sources $S_{4}-S_{21}$, which are located in the 
open waters facing the NEOM coast.   As expected, sources $S_{4}-S_{8}$, which lie closest to the NEOM coastline are characterized by higher impacts 
and shorter arrival times than $S_{9}-S_{21}$.  Overall, sources $S_{4}-S_{21}$ lead to events of various severity, and the histogram 
accordingly exhibits a large scatter over the corresponding segment.  The Duba region appears to be less susceptible to sources lying in the Gulf of Suez,
which are far away from the Duba region, and in the Gulf of Aqaba, except for $S_{35}$ located at the tip of the Gulf which may result with low probability in a large fraction of oil beached near Duba.

Figure~\ref{Duba_prob} illustrates the seasonal distribution of the aggregate probability of volume beached corresponding to oil spills that affect the Duba shoreline. 
As opposed to the Line's shoreline, which is primarily affected by the release sources in The Gulf of Aqaba, sources $S_{4}-S_{12}$ and $S_{14}-S_{15}$ are characterized by the highest aggregate probabilities of volume beached at the Duba shoreline, throughout the year. 
The aggregate probability of $S_{4}$ is the highest in autumn season. 
Few of the sources located in The Gulf of Aqaba are characterized by insignificant probabilities (< 0.01) in the spring ($S_{36}$) and autumn ($S_{36}-S_{37}$) seasons. 
The majority of the sources ($S_{26}-S_{34}$) located farther from the Saudi coastline and closer to Egyptian coast or in the Gulf of Suez are characterized by the lowest probabilities throughout the year.

\subsection*{\textit{Magna, Sharma and Gayal}}

For Magna, Sharma and Gayal, histograms of the volume fractions of oil beached and of risk distributions are shown in Supplementary Figures S7--S8 (Magna), S9--S10 (Sharma) and S11--S12 (Gayal).
For the sake of brevity, the main takeway findings are provided in this section.

% Results for Magna, Sharma and Gayal are analysed and plotted in terms of the histograms of the volume fractions and the distributions of source probabilities, similar to those of The Line and Duba. For the sake of brevity, the plots are shown in supplementary Figures 7 (Magna), 9-10 (Sharma) and 11-12 (Gayal). Therefore, only a few brief remarks are provided in this subsection.

The plots for Magna indicate similarities to those obtained for The Line, where Magna's shoreline is seen to be predominantly  at risk from the release sources in the Gulf of Aqaba.
These sources tend to be associated with the highest impact, with short arrival times and large volumes of oil beached.  Furthermore, the results corresponding to Sharma and Gayal exhibit key similarities with those obtained for Duba. Specifically, the Sharma and Gayal shorelines are primarily vulnerable to the release sources nearest to the Saudi coast, with decreasing risk  from the release sources located far from the Saudi coastline. 
The Gayal shoreline is generally protected from oil spills, which may be attributed to the nearby islands and the shape of its bay. 
In contrast, Sharma's coastline is more exposed to oil spills, where more moderate and high severity events are reported from the release sources lying in the first two rows facing the NEOM shoreline.

\section*{Conclusion}

We conducted a risk assessment associated with accidental oil spills from fixed sources on the NEOM shoreline, focusing in particular on key sites and installations.
For each potential release site, oil spill simulations were conducted over a 28-day period, starting from the first, tenth and twentieth days of each month, over five consecutive years
ranging from 2013 to 2017.  The simulations were carried out using the MOHID's oil spill model, driven with validated high-resolution met-ocean fields of the Red Sea.
The risk associated with each release event was characterized by the minimum travel time for an oil particle to reach the coast, and by the percentage of the total volume of 
oil released that was beached on the NEOM coast at the end of the simulation period. 

The results indicate that spills originating in the Gulf of Aqaba are characterized by short arrival times and high volume fractions, making them the most hazardous to the NEOM shoreline. 
This occurs throughout the year except for the summer months, when the prevailing southwards currents in the Gulf of Aqaba tend to push the oil slicks towards the Tiran and Sanfir 
islands, which does not minimize their potential impact because these islands are key sites for tourism.  Release sources located in the open water closest to the Saudi Arabian shoreline are generally associated with short arrival times, except during the months of September and October.  These release sources impact NEOM's islands and the region connecting Sharma to Duba throughout the year.  On the other hand, these release sources have weak impact on the NEOM shoreline lying in the Gulf of Aqaba, between June and December.
Release sources located in the Gulf of Suez have a slight impact on the NEOM shoreline during the months of January, Februrary and March.  Finally, spills originating from release sources located in the open waters close to the Egyptian coast are characterised by moderate arrival times and low volume fractions, throughout the year.

The shorelines of Magna and The Line are subject to a similar response to the oil spill scenarios considered, where both were vulnerable to the release sources located in the Gulf of Aqaba.
Moreover, release events south of Tiran and near Sanafir islands may have a significant impact on The Line's shore, particularly during winter and more so in spring. 
Duba, Sharma and Gayal's shorelines exhibit similar behavior in response to accidental oil spills from the sources considered.
Specifically, release sources lying closest to the Saudi Arabian shoreline have the biggest impact on the shorelines of these sites. 
The releases are characterized by short arrival times and large fractions of volume beached.  The adjacent release sources also exhibit a considerable impact, 
that is weaker during the Autumn months.  These release events are typically associated with medium severity arrival times and fractions of volume beached. 
Finally, Duba, Sharma and Gayal's shorelines appear to be at low risk from accidental oil spill scenarios originating from release sources near the African shoreline 
during the summer and autumn seasons.

\bibliography{sample}

\section*{Acknowledgements}
The work was funded by the Office of Sponsored Research (OSR) at King Abdullah University of Science and Technology (KAUST) under the Virtual Red Sea Initiative (Award No. REP/1/3268-01-01), OSR-Competitive Research Grant (Award No. OSR-CRG2018-3711), and the Saudi ARAMCO-KAUST Marine Environmental Research Center (SAKMERC). The simulations were performed on the KAUST supercomputing facility, SHAHEEN.

\section*{Author contributions statement}
H.V.R.M. and M. A. E. H. conceptualization, methodology, software, data curation, formal analysis, writing original draft, A. K. C. conceptualization, methodology; I.H. and O.K. conceptualization, funding acquisition, resources, writing-review and editing.

\section*{Competing interests}
The authors declare that they have no known competing financial interests or personal relationships that could have appeared to influence the work reported in this paper.

\section*{Data Availability}
The datasets used and/or analyzed during the current study are available from the corresponding author upon reasonable request.

\clearpage

\clearpage

\clearpage

% \clearpage

\clearpage

\clearpage

\clearpage

\clearpage

\end{document}